\documentclass{article}%
\usepackage{amsmath}
\usepackage{amsfonts}
\usepackage{amssymb}
\usepackage{graphicx}%
\providecommand{\U}[1]{\protect\rule{.1in}{.1in}}
\newtheorem{theorem}{Theorem}

\newtheorem{definition}[theorem]{Definition}

\begin{document}

\title{The Einstein-Jordan conundrum and its relation to ongoing foundational
research in local quantum physics \\{\small Dedicated to the memory of Claudio D'Antoni}\\{\small to be published in EPJH}}
\author{Bert Schroer\\present address: CBPF, Rua Dr. Xavier Sigaud 150, \\22290-180 Rio de Janeiro, Brazil\\email schroer@cbpf.br\\permanent address: Institut f\"{u}r Theoretische Physik\\FU-Berlin, Arnimallee 14, 14195 Berlin, Germany}
\date{November 2012}
\maketitle

\begin{abstract}
We demonstrate the extraordinary modernity of the 1924/25 "Einstein-Jordan
fluctuation conundrum", a Gedankenexperiment which led Jordan to his
quantization of waves published as a separate section in the famous
Born-Heisenberg-Jordan 1926 "Dreim\"{a}nnerarbeit". The thermal nature of
energy fluctuations caused by the restriction of the QFT vacuum to a subvolume
remained unnoticed mainly because it is not present in QM.  In order to
understand the analogy with Einstein's fluctuation calculation in a thermal
black body system, it is important to expose the mechanism which causes a
global vacuum state to become impure on a localized subalgebra of QFT. 

The present work presents the fascinating history behind this problem which
culminated in the more recent perception that "causal localization" leads to
thermal manifestations. The most appropriate concept which places this
property of QFT into the forefront is "modular localization". These new
developments in QFT led to a new access to the existence problem for
interacting quantum fields whose solution has remained outside the range of
renormalized perturbation theory. It also clarifies open problems about the
relation of particles and fields in particular about the incompletely
understood crossing property. Last not least it leads to a constructive
understanding of integrable versus non-integrable QFTs..

\end{abstract}

\section{QFT, how it begun and how modern concepts solve the Einstein-Jordan
conundrum}

It is well known that, long before the observational discovery of the photon,
Einstein postulated a corpuscular nature of light \cite{E1} based on thermal
fluctuation properties of black-body radiation. In a detailed theoretical
analysis of subvolume fluctuations in a semiclassical (Bohr-Sommerfeld)
statistical mechanics \cite{E2}, he identified, in addition to the expected
wave-like fluctuation component, a particle-like component which he
interpreted as an indication of a corpuscular aspect of light. In his view the
presence of this component was important for attaining thermodynamic
equilibrium. Whereas the photoelectric effect constituted the first
observational support for photons, Einstein's 1917 fluctuation result
\cite{E2} which confirmed his earlier ideas \cite{E1} was of a purely
theoretical kind. In view of the inaccessible nature of such fluctuations to
direct observations, Einstein's argument remained a famous Gedankenexperiment
in the setting of the old quantum theory.

This work attracted the attention of one of Max Born's younger collaborators
who had some familiarity with statistical mechanics and first hand knowledge
about the newly developing quantum theory. In his 1924 Ph.D. thesis
\cite{Thesis} Pascual Jordan contradicted Einstein's assertion that one needs
the presence of a particle-like component (Nadelstrahlung) to obtain
thermodynamic equilibrium. Einstein's answer came swiftly; only some month
later he published a counter paper \cite{E3} in which he showed that, despite
the confirmed mathematical correctness of Jordan's thesis, there was a problem
of a more physical nature on which he failed, namely the particle-like
component was needed in order to get the right absorption coefficients.

This encounter with Einstein, which Jordan entered as an adversary of the
theory of (what later was called) photons, did not only ruffle his feathers
but, fortunate for the birth of particle theory, also put him onto his
figurative "road to Damascus" in that he became the discoverer of quantum
field theory (QFT), and the most uncompromising enunciator of a quantum theory
(QT) of light \textit{and} (de Broglie wave) matter within a unified setting
of quantized fields.

There is a subtle irony in the fact that Jordan's radical change of mind was
caused by Einstein, a lifelong opponent of quantum probabilities. If Jordan's
claim of a complete analogy of quantum fluctuations in a subvolume with
Einstein's thermal fluctuations really amounted to attribute a thermal aspect
to his new wave quantization, then the new theory in its restriction to a
subvolume should also admit an intrinsic probability namely that of thermal
ensembles, with which Einstein was entirely familiar.

It can be assumed that neither Einstein nor Jordan were aware of these
implication of the E-J "conundrum" \cite{Du-Ja}. As a result Jordan, in his
dispute with Born and Heisenberg, did not receive the support of Einstein
which he had hoped for. Looking back at this episode with historical
hindsight, a great chance, which may have steered QFT from its very beginnings
into a more foundational direction, was lost. Born's addition of a probability
concept to events caused by global observables\footnote{QM is a global quantum
theory; there is no localization which leads to subalgebras (ensembles) of
observables. Whereas the ensemble point of view in QFT and its
thermo-probabilistic manifestantions is intrinsic, in QM it is up to the
interpreter. This freedom has led to many heated disputes.} and states in
quantum mechanics (QM), and some of the counter-intuitive aspects of QT would
perhaps have appeared in a different light. Although there is no probability
of thermal origin in QM, it is the more fundamental QFT which has the
suzerainty over the conceptual problem about the nature of probability. In
fact the main aim of the present work is to convince the reader that the
concept of ensembles, which is inherent in Haag's presentation of QFT in terms
of (subvolume) localized algebras, is the best setting for a complete
resolution of the E-J conundrum.

In the famous "Dreim\"{a}nnerarbeit" with Born and Heisenberg \cite{BHJ},
Jordan contributed a separate section containing a calculation of the mean
square energy fluctuations in a subvolume for a simple model of quantized
waves (2-dimensional "photons") applied to a two-dimensional free wave
equation (see section 4); a model which he considered as a one-dimensional
analog of Maxwell's theory of light. The result of his approximate
calculation\footnote{Even though the system is noninteracting, subvolume
fluctuations do not permit a computation in closed form.} consisted in the
verification of the presence of wave and particle components in the subvolume
fluctuation spectrum, just as in Einstein's statistical mechanics calculation,
except that Jordan's global state was the vacuum state of QFT, whereas
Einstein was analyzing subvolume fluctuations in a global thermal
("heat-bath") state in statistical mechanics. Jordan did not address the
problem how subvolume (subinterval) fluctuations of a quantized system in a
ground state can mimic those of a global thermal system; it is not even clear
whether he realized that this is not possible in QM.

In the setting of Schr\"{o}dinger quantum mechanics one knows that the global
vacuum tensor-factorizes into the vacuum of a spatial subsystem and its
complement\footnote{In order to facilitate a comparison of QM with QFT we will
use throughout this article the "second quantized" form of QM for which the
Schr\"{o}dinger wave function is replaced by a quantum field.}. Can the vacuum
of QFT deviate from this quantum mechanical behavior i.e. could the local
restriction of a global vacuum lead to thermal behavior? As a result of these
kind of conceptual questions, which remained open in Jordan's contribution and
were forgotten afterwards when QFT was on its path to success, the name
\textit{Einstein-Jordan conundrum} \cite{Du-Ja} is quite appropriate. In the
present work it will be shown that the missing thermal properties of subvolume
fluctuation in a global vacuum state of QFT can be verified and hence the
conundrum aspect disappears. In Jordan's model of a two-dimensional wave
equation, the analogy of subvolume-localized QFT with the thermal aspects of
statistical mechanics amounts to an isomorphism of a global heat bath thermal
system with a localization-caused thermal aspect of QFT (section 4).

Jordan's quantum theoretical calculation, which appeared as one section in the
Dreim\"{a}nnerarbeit \cite{BHJ}, did not receive the unrestricted endorsement
of his coauthors Max Born and Werner Heisenberg \cite{Du-Ja}; too many
assumptions in Jordan's treatment of infinities and other unclear aspects in
his approximation were swept underneath the rug and prevents them from
embracing Jordan's remarkable but somewhat suspicious calculation. Whereas the
calculation techniques of the new quantum mechanics were usually transparent,
this was not the case in Jordan's field theoretical model.

Nowadays we know good reasons for such doubts; their conceptual origin is the
clash between the \textit{causal localization} of relativistically propagating
QFTs and the much simpler localization structure of QM\footnote{Intuitively
speaking this is the difference between infinite velocity propagation (action
at a distance) and propagation with a limiting velocity. QM, whether
formulated a la Schroedinger or second quantized, has no limiting velocity in
its algebraic structure and any finite velocity as that of sound is
"effective" i.e. arises in suitable expectation values for large times.} based
on the \textit{Born localization} (resulting from the spectral theory of the
quantum mechanical position operator). Whenever one has to rely on
approximations, as in subvolume fluctuations of QFT, one must verify their
compliance with causal localization.

It is well known that perturbative QFT led to serious problems with the
formalism of QM for a long time until it was formulated in a
\textit{relativistic covariant }way (closely related to causal localization)
at the end of the 40s. In fact the clearest formulation of relativistic
perturbative theory results from the iterative implementation of the causality
principle in the form of spacelike commutations of fields, which is known as
the Epstein-Glaser approach \cite{E-G}. In other formulations of perturbation
theory using regularizations and cut-offs (which are often computationally
more efficient) this is less clear. In addition there is the human aspect of
forgetting the principles behind a formalism once it has been formulated in
terms of efficient computational recipes.

In the case of the subvolume fluctuation problem the elegance of relativistic
covariant perturbation theory is of no help; it is better to verify the
consistency of an approximate calculation by using the closely related
\textit{causal localization} in a more direct manner. One of its
manifestations is the use of the thermal KMS\footnote{The
Kubo-Martin-Schwinger analytic characterization of thermodynamic limit states
replaces the tracial Gibbs state (density matrix) formalism which breakes down
as a result of volume divergence for V$\rightarrow\infty.$} \cite{Haag}
structure of the restricted vacuum state and to approximate this rather
singular state by Gibbs density matrix states. Such problems are at best
formulated and solved in the new \textit{modular localization setting}
(section 3).

Although there is a formal analogy to the thermodynamic limit in the heat bath
(statistical mechanics of open systems) setting, the Hamiltonians of modular
theory are generally not those which correspond to time translations of a
non-inertial observers in Minkowski spacetime; in fact the automorphism of the
localized algebra which they generate has generally no interpretation at all
in terms of a \textit{geometric} flow within the causally extended
localization region (\textit{fuzzy} automorphism). Beyond its preservation of
the localization region, almost nothing is known for compact causal
localization regions.

The theory which describes such fluctuation phenomena in a model-independent
way is a special case of an abstract mathematical theory of operator algebras,
which carries the name of its protagonists: the \textit{Tomita-Takesaki
modular operator theory} \cite{Haag}. In quantum physics one encounters this
theory in two places: statistical mechanics (in particular in the Gibbs
formulation and its thermodynamic limit), and in localization problems of QFT
(such as that of the E-J conundrum). In this second role the setting is often
referred to as \textit{modular localization }\cite{S2}. Whereas the
statistical quantum theory of open systems is most elegantly formulated in the
setting T-T modular theory, the latter becomes really indispensable in the
context of modular localization. It is the only way to describe the
model-independent thermal nature of spacetime-restricted vacuum states in
local quantum physics (LQP). Since a mathematically rigorous presentation of
this setting would go beyond what one can reasonably expect of a reader with
interest in the conceptual aspects of QFT to digest, we will sacrifice
mathematical precision in favor of conceptual physical understanding. In this
introduction and the next section some of the concepts will still retain their
intuitive metaphoric meaning, only in the subsequent section some
mathematical/conceptual precision will be added.

Even though a free quantum field obeying a linear wave equation can be
adequately described in terms of global quantum mechanical
oscillators\footnote{The claim that a free field theory associated with
particles of arbitraey spin "is nothing else that a collection of oscillators"
is an exaggeration since a student of quantum mechanical oscillators would not
be able to combine them into a covariant field.} (momentum space
creation/annihilation operators), this quantum mechanical description is not
useful for QFT fluctuation problems in subvolumes. As mentioned the
localization-induced thermal aspects of the E-J conundrum, which any
approximate calculation should fulfill, are hard to reconcile with global
quantum mechanical oscillator descriptions. More specifically, it is not clear
how the global oscillators, in terms of which a free field can be written, can
retain their utility in subvolume problems.

Spatial localization of quantum mechanical variable in terms of projectors
associated with spatial regions which appear in the spectral decomposition of
the position operator ("Born-localization") lead (after applying them to
global states or operators) to Born-localized states or to local observables
at a fixed time; the most convenient way to see this is a Fock space
formulation of QM ("second quantization") which maintains the physical
content, but brings QM into a formal analogy with QFT. In causal QFT such a
spatially localized algebra is equal to the algebra localized in its spacetime
causal completion which in the simplest case of a spatial ball is the double
cone extended by the "causal shadow" with the ball as its base.

Whereas in QM the $\mathbf{x}$ ranges through the spectrum of the position
operator, the points of Minkowski spacetime $\mathbf{x,}t,$ which parametrize
relativistic fields, have no such operator interpretation. Hence the quantum
mechanical localization is directly linked to the probability interpretation
which Born \cite{Born} added to Heisenberg's QM shortly after the
Dreim\"{a}nnerarbeit. As a result of absence of a position operator in
QFT\footnote{QFT in this article always refers to relativistic QFT.}, the Born
probability looses its algebraic realization in terms of observables and
continues to be important for wave functions\footnote{Even in QM its use for
wave functions is physically more important than for second quantized
Schr\"{o}dinger fields. Propagation velocities (e.g. the velocity of sound)
describe the asymptotic movements of the position of maximal probability
density of wave functions. Despite the frame dependence of Born localization
its large time consequence for the movement of centers of relativistic wave
packets in QFT comply with independence on inertiel frames.} (vector states).

The difference in localization leads to significantly different mathematical
structures and physical consequences. Algebras at equal times in the
\textit{Fock space }formulation of QM tensor-factorize into an $\mathcal{O}%
$-localized subalgebra and that localized in its spatial complement
$\mathcal{O}^{\prime}.$ Operator algebras in QFT do not share this property,
even though both algebras commute and together generate the global algebra.
Related to this is the property of factorization of the nonrelativistic
vacuum, whereas the QFT vacuum under subdivision becomes entangled in a very
strong (singular KMS) sense associated with infinite vacuum polarization
"clouds" at the causal boundary \cite{interface}. In fact local operator
subalgebras in QM are with respect to their von Neumann type the same as their
global counterpart namely isomorphic to $B(H),~$the operator algebra of all
bounded operators on a Hilbert space $H$, and the Hilbert space suffers an
inside/outside factorization $H=H_{inside}\otimes H_{outside}$ which follows
the spatial split.

Local algebras in QFT are radically different, they are all isomorphic to an
operator algebra which, for reasons which become clear later on, will be
referred to as a \textit{monad}. Saying that they act in a Hilbert space $H,$
and therefore are subalgebras of $B(H),$ does neither help to understand their
mathematical properties nor their physical role. In contrast to QM, a
halfspace algebra at a fixed time (or its associated causally completed
wedge-locaized algebra) and its opposite halfspace counterpart (causally
disjoint wedge) commute but do not tensor-factorize (monads do not factorizes
with their commutants). This leaves room for a very singular kind of
entanglement which cannot be described in the standard setting of quantum
information theory \cite{integrable}. This kind of entanglement does not have
to (and should not) be averaged over the "opposite" degrees of freedom; the
associated probabilistic KMS state ("singular density matrix") is obtained
just by subalgebra-restriction of the original global vacuum. 

We promised the reader to refrain from damping his/her interest in conceptual
historical problems reaching back to the dawn of QFT by presenting technical
mathematical details. The only exception will be those cases for which
technicalities admit a simple physical interpretation. One such case is that
operator algebras in the context of LQP weakly closed i.e. they are von
Neumann algebras. As the result of their algebraic characterization as
consisting of subalgebras of $B(H)$ which remain preserved under the two-times
application of forming commutants in $B(H),$ as well as the fact that
commutants play an important role in the formulation of Einstein causality
(statistical independence of spacelike separated measurements), the use of
such algebras enjoys direct physical support. The local subalgebras of QFT are
always factors (indecomposability).

Another difference from QM is the use of the word state and (state) vector.
States are positive linear functionals $\omega$ on operator algebra i,e,
$\omega(A),~A\in\mathcal{A}.$ Their physical meaning and the problem of their
representation in terms of vectors in a Hilbert space as $\omega(A)=\left(
\psi,A\psi\right)  ,~\psi\in H,~A\in\mathcal{A}$ depend on the structure of
the algebra. In QM where $\mathcal{A}=B(H),$ independent of whether
$\mathcal{A}$ denotes a global algebra or a Born-localized subalgebra (in
which case $H~$is a subspece of the total Hilbert space), the state determines
a vector uniquely up to a phase factor and therefore the identification of
states with vectors in $H$ makes good sense. This kind of uniqueness breaks
down for other operator algebras which appear in the classification of
factors, in particular for states on a monad. In cyclic representations
(existence of a vector on which the application of the algebra creates a dense
set in a Hilbert space) there is a unique relation between the algebra and a
dense set of vectors, but the representation of states by vectors remains
highly non-unique. The distinction between states and vectors is crucial in
the study of localized subalgebras of QFT which are always of the algebraic
monad structure.

This immense structural difference resulting from Born-localization in QM as
compared to modular localization in QFT has been generally overlooked outside
of LQP; in fact the latter may be understood as a formulation of QFT which
highlights precisely these differences. An educated guess why QFT developed in
this way is that QM and QFT share the formalism of Lagrangian quantization and
functional integral representation and the important renormalized perturbation
theory is based on computational recipes which do not place the antagonism of
the underlying localization principles into sufficient evidence.

Whereas the functional integral approach is a rigorous mathematical tool in
QM\footnote{Even in QM it is not advisable to use functional integrals in a
course on QM where exact solutions (integrability) are presented.}, the lack
of its mathematical control in QFT is partially compensated by its intuitive
suggestive content which together with some corrective hindsight often leads
to correct recipes for perturbative calculations. As a result of the well
established divergence of perturbative series, such calculations do not say
anything about the existence of a model of QFT; but perturbation theory comes
with a lower level of consistency (that of formal power series) which
facilitates its extraction from mathematically nonexistent functional
integrals with a modest amount of hindsight; it is not sufficient to show that
the perturbative result admits a functional integral representation. Questions
as to why in important cases the low terms of diverging power series lead to
incredibly precise agreement with experimental data are not really answered by
claiming that these power-series are asymptotically convergent in the limit of
vanishing interaction strengths; as long as the existence of a model remains
unproven such claims have no mathematical basis.

The problem of approximating the subvolume energy fluctuations in QFT is quite
different since, at least in the absence of interactions, the existence is
secured. However such problems cannot be solved without resorting to
approximations, so the remaining task is to show that such approximations
remain compatible with localization and its thermal aspects. In this respect
the improved calculation in the Jordan model proposed in the work of Duncan
and Janssen is somewhat contradictory because by claiming that the restricted
vacuum remains a pure state (see remarks after equ. (53) in \cite{Du-Ja}), one
throws away the child with the bath tub\footnote{Since this is a common
conceptual misunderstanding of QFT, the criticism is not personal; in fact it
is probably the way Jordan considered his subvolume fluctuations.}. Thinking
in terms of QM on the other hand, it is natural to add a coupling to an
external heat bath in order to enforce the thermal aspect (related to their
belief that the subinterval restriction does not effect the purity of the
restricted state) of the conundrum and this is precisely what Duncan and
Jannsen did; but perhaps the vacuum really does not factorize in their
approximation in which case there would only be a discrepancy between their
(and Jordan's) possibly correct approximation with an incorrect verbal claim
about factorization (which negates the thermal impure nature).

Often physicists use loose language by calling QFT "(relativistic) quantum
mechanics". This incorrect terminology can create conceptual havoc in those
cases in which the emphasis on the differences becomes essential. To remain
clear on this point, it may be useful to mention that relativistic quantum
mechanics as being something different from QFT really exists; it is known
under the name \textit{direct particle interactions} \cite{Coe} and describes
a theory which is solely formulated in terms of particles and their
Poincar\'{e} invariant scattering matrix (no covariant local observables) and
has no conceptual relation with QFT \cite{interface}.

Many articles and books create the impression that the mere existence of
infinite degrees of freedom separates QFT from QFT. But as the existence of a
second quantized presentation of QM shows, this is not the case. What is
however true is that in QM the appropriately defined phase space density
(degrees of freedom per unit cell of phase space) is finite, whereas the
causal localization of QFT requires a "mildly" infinite ("nuclear") phase
space degree of freedom behavior \cite{Haag}.

It is very difficult to check by hand within the standard setting of QFT
whether a calculation is consistent with thermal aspects of subvolume
fluctuations; it is easier to use a formulation of QFT which takes the thermal
aspect into account from the outset. An adequate setting which guaranties that
thermal aspects of localization are correctly implemented can be given in the
setting of \textit{local quantum physics} (LQP), also referred as
\textit{algebraic quantum field theory }(AQFT); such a setting dates back to a
seminal 1957 talk by Haag (for a recent translation see \cite{Lille}). From
its humble beginnings it has developed into a nonperturbative mathematically
precise intrinsic setting of QFT i.e. a formulation of QFT which does not
depend on a quantization parallelism to classical field theory (last section);
in particular it does not require to understand how thermal aspects, which are
absent on the classical level, emerge through quantization.

The conceptual progress of QFT has revealed that Haag's intuitive idea of
localization in terms of local observables, envisaged as counters which have a
finite extension in space and are switched on for a finite duration in time,
leads to unexpected somewhat metaphoric situations if its exact mathematical
formulation is re-interpreted in terms of a Gedankenexperiment. Even in the
simplest of all cases, the noncompact localization in Rindler wedges of
Minkowski spacetime, the appearance of a measurable thermal radiation in the
Unruh Gedankenexperiment (see later) requires to uniformly accelerate the
hardware with absurdly big acceleration which cannot be achieved with
macroscopic counters. Such Gedankenexperiments reveal an unexpected side of
localization. They focus attention to an aspect of QFT which, although
somewhere hidden in the Lagrangian quantization setting, is naturally
accounted for in the LQP formulation of QFT.

There is hardly any conceptual enrichment of QFT which has been as fruitful
for this kind of problems as Haag's algebraic LQP setting which describes the
model independent nature of such phenomena. Direct attempts at physical
realizations of principles in form of Gedankenexperiments may acquire a
somewhat metaphoric counter-intuitive appearance (perhaps the reason why the
Unruh effect has led to many controversies), but as long as their mathematical
formulation is precise and sufficiently many (possibly indirect) physical
consequences agree with observational tests a theory is successful.

The strategy underlying the LQP setting is in a way opposite to that of
quantization which is based on analogies to classical physics which, with the
exception of QM remain mathematically vague and whose underlying physical
principles cannot easily be seen from the computed perturbative results. On
the other hand in LQP one first formulates the principles and properties which
a physically acceptable QFT should have in a mathematically rigorous way; only
afterwards one looks for methods to classify and construct models which
fulfill these "axioms" and comply with experimental observations
(top-to-bottom approach).

It is the main aim of the present work to explain these properties and show
how the Einstein-Jordan conundrum, including its thermal aspects, can be
understood. For Jordan's model of a two-dimensional zero mass wave equation
the relation is explicit and amounts to an isomorphism of the two systems
(section 4). The foundational aspects of QFT were present since its beginnings
in 1925, but their understanding in the ongoing research is still a project
which, different to QM, had yet not arrived at a conceptual closure.

Jordan's coauthors Born and Heisenberg felt that his presentation of subvolume
fluctuations of quantized waves did not quite fit into the quantum mechanical
setting of their joint work, which was the second paper written after
Heisenberg's presentation of QM. They did not see that Jordan with his
quantization of waves discovered the beginnings of a new QT which went beyond
the scope of QM. In a letter Heisenberg challenged Jordan several years later
\cite{Du-Ja} to account for a term which diverges proportional to
$log\varepsilon^{-1}$ and which Jordan should have seen in his two-dimensional
model of quantized waves; here $\varepsilon$ is a measure of "fuzzyness" at
the endpoints of the localization interval. This shows that the differences to
QM of Jordan's wave quantization was slowly being appreciated.

This correspondence preceded Heisenberg's famous paper on vacuum polarization
\cite{1934}; it represents the diverging contribution from vacuum polarization
in the limit of sharp localization. In his paper Heisenberg implicitly
proposed that for models in 4-dimensional spacetime vacuum polarization caused
by localizing dimensionless observables leads to a divergence proportional to
the dimensionless "fuzzy" surface $A/\varepsilon^{2}.$ He exemplified this in
the case of a dimensionless \textit{partial} charge localized in a finite
volume and showed that the formation of particle/antiparticle pairs near the
surface is the cause of this behavior. It has no counterpart in classical
field theory and in QM. The distribution theoretical setting which permits a
rigorous derivation of this behavior of partial charges from the singular
properties of their pointlike conserved currents became available only after
it was realized that relativistic covariant fields and currents were
operator-valued distributions \cite{Wight} i.e. objects which had to be
smeared with test functions before they became (generally unbounded)
operators. This will be briefly sketched in the next section.

Heisenberg's vacuum fluctuations are closely related to the thermal aspects of
localization, but it is not so easy to see this in the standard quantization
formalism. The setting of Haag's LQP, in which the concept of modular
localization of states and operators permits a natural and precise formulation
is therefore better suited for this purpose. It permits to explore the
powerful Tomita-Takesaki modular operator theory for the localization problems
of QFT.

The first step in linking QFT with the T-T modular theory was the realization
that the \textit{statistical mechanics of open systems}, i.e. the direct
description of thermal states in the infinite volume limit, is a special case
of the T-T modular theory. The limiting states loose their characterization as
density matrix Gibbs states; what remains is their KMS property which, before
it became the defining property of thermodynamic limit states, was used in the
work of Kubo, Martin and Schwinger as an analytic tool which facilitated the
calculation of traces of Gibbs density matrices. In a seminal paper by Haag,
Hugenholtz and Winnink published in 1967, this observation was elevated to a
fundamental property which follows from the stability requirements of
thermodynamic equilibrium \cite{Haag}.

It was a lucky coincidence that physicists working with operator algebraic
methods met rather early (at the 1967 international conference in Baton Rouge,
see \cite{Bo}) with mathematicians who already had obtained important results
on operator algebras which generalized what had been obtained from the study
of (unimodular) Haar measures within group representation theory and became
referred to as "modular operator theory". In this way both theories were
combined; the mathematicians incorporated the ideas around the conceptual use
of KMS and the physicists adopted the Tomita S-operator and the
(Tomita-Takesaki) modular theory.

The important point in the present context is that the algebra changed its
nature in the thermodynamic limit; whereas the approximating box-quantized
statistic mechanic algebras are of type I\footnote{This is the standard
algebra encountered in QM consisting of all bounded operators B(H) of a
Hilbert space H.} i.e. isomorphic to the $B(H)$ global algebras (both in
QM/QFT), the open system (thermodynamic limit) algebra is (in QM and QFT) a
hyperfinite factor algebra of type III$_{1}$ in Connes extension of the
Murray-von Neumann classification. In fact the KMS property provided by
modular operator theory played an important role in Connes refinement of
classification of factor algebras \cite{Co}. More recently there were attempts
to interpret time as originating from the operator algebraic KMS property
\cite{Con} which is opposite to relating proper time with (the Unruh) temperature.

The property which secures the applicability of that theory is the
\textit{standardness} of the thermal operator algebra i.e. that the existence
of a vector $\Omega$ representing a thermal state in the Hilbert space $H$ on
which the application of the operators $A$ of the algebra $\mathcal{A}$
generate a dense set of states (cyclicity) and does not contain nontrivial
annihilation operators:%
\begin{align}
&  the~subspace~\mathcal{A}\Omega\text{ }is\text{ }dense~in~H~(cyclic)\\
if~A\Omega &  =0~for~A\in\mathcal{A\subset}B(H)\curvearrowright
A=0~(separating)\nonumber
\end{align}
where $B(\mathcal{H})$ denotes the algebra of all bounded operators. The T-T
operator theory associates with such a standard pair a densely defined
involutive antilinear Tomita S-operator%
\begin{align}
SA\Omega &  \equiv A^{\ast}\Omega,~S^{2}=1~on~its~domain\text{ }%
domS\label{S}\\
S  &  =J\Delta^{\frac{1}{2}}=\Delta^{-\frac{1}{2}}J,~\nonumber
\end{align}
for which its closure (denoted with the same letter) has the polar
decomposition (written in the second line) in terms of antiunitary reflection
$J$ and a positive operator $\Delta$ which leads to the unitary modular group
$\Delta^{it}.$ Whereas the proof of the above properties is rather
straightforward, proving the T-T theorem is anything but simple
\cite{Robinson}. It states that $J$ maps $\mathcal{A}$ into its algebraic
commutant $\mathcal{A}^{\prime}$ (the algebra of operators in $B(H)$ which
commute with every operator in $\mathcal{A)},\mathcal{~}$and $\Delta^{it}$
defines a modular automorphism $\sigma_{t}$ of $\mathcal{A}$
\begin{align}
J\mathcal{A}J  &  =\mathcal{A}^{\prime},~\sigma_{t}(A)=\Delta^{it}%
A\Delta^{-it}\in\mathcal{A~}for~A\in\mathcal{A},~\Delta^{it}=e^{-itH_{mod}%
}\label{T-T}\\
\omega(A)  &  \equiv\left(  \Omega,A\Omega\right)  ,~KMS:~\omega(A_{1}%
A_{2})=\omega(A_{2}e^{-H_{mod}}A_{1}),~A_{i}\in\mathcal{A}~\nonumber
\end{align}
where the introduction of the modular Hamiltonian serves to show that the
modular KMS in modular theory the dimensionless "temperature" corresponds to
$\beta=1.~$For more details of the description of equilibrium statistical
mechanics in this setting the reader is referred to \cite{Haag}.

The presentation of the subsequent sections is facilitated by adding some more
notation and comments on its physical meaning. Although \textit{global}
algebras in ground state problems (in contrast to thermal states) of both QM
and QFT lead to the same type of algebras of all bounded operators in a
Hilbert space $B(H)$, there can be nothing more different than the result of
passing to localized subsystem. In QM the spatially localized subalgebras
remain $B(H_{sub})$ algebras where $H_{sub}=PH$~with $P$ a projector from the
spectral decomposition of $\mathbf{x}$ associated with that part of the
spectrum which corresponds to the localization region $\mathcal{C}$
\begin{align}
&  nonrel.:~B(\mathcal{C})=B(H(\mathcal{C})),~type~I_{\infty}\\
QFT  &  :\mathcal{A(O}),~hyperfinite~type~III_{1}="monad"\nonumber
\end{align}
whereas in the QFT case it undergoes a radical change in that all the
localized algebras are of the type of a monad independent of the localization
region. "Monad" is a short hand terminology for an isomorphy class of
indecomposable representations of a unique von Neumann factor algebra
(hyperfinite type III$_{1}$ factor algebra) where the factor property replaces
the irreducibility of $B(H).$ The timelike causal shadow property (see
(\ref{sha}) next section) permits to replace a simply connected spatial
localization $\mathcal{C}$ by the spacetime localization in the associated
causal shadow $\mathcal{C\rightarrow C}^{\prime\prime}.$ Algebras which result
from sharp localization in QFT are always isomorphic to a monad; only suitably
(\textit{split-property }\cite{Do-Lo}) defined "fuzzy"-localized algebras can
be of type $B(H(\mathcal{O}_{f.l.})$ with $\mathcal{O}_{f.l.}$ the spacetime
region of fuzzy localization.

By using the terminology "monad" we do not expect the reader to know the
classification of operator algebras; for a physicist it is more important to
understand a monad through its physical properties \cite{Jakob}; some of them
will appear in the next three sections. Sharp causal localization cannot be
expressed in terms of projectors; localization in terms of projectors and
(sharp) causal localization are mutually exclusive. The most surprising if not
spectacular physical property of a monad is that a full QFT (including its
Poincar\'{e} group acting on a d-dimensional Minkowski spacetime) can be
encoded into an abstract Hilbert space positioning of a finite number of
copies of a monad without any internal structure. This attributes to QFT an
ultra-relational aspect which no other type von Neumann factor is capable of
generating (see later).

Although some physicists with an early familiarity with LQP probably knew
that, as a result of the Reeh-Schlieder property of localized subalgebras in
QFT the standardness of a spacetime localized algebra $\mathcal{A(O})$ with
respect to the vacuum and therefore the applicability of the T-T modular
theory always holds, the awareness about its physical implications had to wait
another decade. It came in a paper by Bisognano and Wichmann \cite{Bi-Wi} when
these authors realized that the above modular objects in the case of wedge
algebras have a geometric physical interpretation: the unitary modular group
is given in terms of the operators representing the wedge-preserving Lorentz
boosts and the $J$ is the antiunitary operator which reflects on the edge of
the wedge (the TCP operator up to a $\pi$-rotation). A special $z-t$ wedge
region is defined as $W_{z,t}=\left\{  z>\left\vert t\right\vert
;~x,y\in\mathbb{R}^{2}\right\}  $ and a general wedge is obtained by applying
Poincar\'{e} tranformations.

The proof given by B-W, which leads to the mentioned geometric identification
of the modular objects, still depended on some reasonable technical assumption
about operator algebra properties resulting from quantum fields. A direct
proof in the algebraic setting of LQP can be found in \cite{Mu}; this proof is
free of technical assumptions and uses besides the requirements of LQP the
physically motivated property of the validity of a complete particle
interpretation. In this way the modular wedge localization of Wigner wave
functions \cite{BGL} turns out to be useful for proving its interacting
algebraic counterpart.

In general the modular objects do not act geometrically in other cases, but
the modular unitaries always respect the causal boundaries of localization and
the local algebras can be defined (if necessary by "dualization") in such a
way that the commutant is equal to the algebra associated to the causal
complement (Haag duality \cite{Haag}). The modular unitaries of other regions
$\mathcal{O}=\cap_{W\supset\mathcal{O}}W$ which allow a geometric
representation in terms of intersections of wedges, the modular data of
$\mathcal{A(O)}$ are geometric in the indirect sense of being determined in
terms of the modular data of the participating intersecting wedge algebras.

Since the unitary modular groups of wedges are the unitaries of the
wedge-preserving Lorentz boosts and the representation of the Poincar\'{e}
group is shared between the interacting theory and its free field asymptote,
the only dependence on the interaction can be in $J.$ Indeed the interacting
$J$ is connected with its free field counterpart $J_{0}$ via the scattering
matrix $J=S_{scat}J_{0}$ \cite{S2}. The conceptual simplicity of the
definitions stands however in stark contrast to the difficulties one
encounters in attempts to calculate the modular data of intersections.

This concept of modular localization leads to thermal manifestations of
localization in terms of modular KMS properties. They share some properties
with statistical mechanics (localization-caused versus heat bath thermal
behavior). A pure global vacuum state reduced to a spacetime subregion becomes
a highly entangled state with a rather singular notion of entanglement which
is shared by all monad representation independent of their origin. The sharply
localized reduced state $\omega$ cannot be described in terms of a density
matrix inasmuch as a density matrix Gibbs state looses this property in the
thermodynamic limit. Despite the shared KMS property between the "heat bath
thermality" of statistical mechanics and the "localization thermality" of QFT
there are also important differences. Localization thermality is more abstract
because the modular Hamiltonian is never related to the translation in
Minkowski spacetime, it is rather intrinsically determined by the standard
pair $(\mathcal{A(O)},\Omega)$.

As a consequence the modular analog of temperature cannot be directly
measured, its main physical purpose is to describe the singular (i.e. not
density matrices) impurity of subvolume-reduced vacua. In the last section
this theory will be applied to Jordan's model. This belongs to the class of
conformal QFTs which have admit geometric modular groups for certain compact
regions which arise by applying conformal maps to wedges. In the case of
Jordan's chiral conformal model the relation between an Einstein statistical
mechanics and the restricted vacuum QFT is an isomorphism. Such special
situation have been referred to as an "inverse Unruh effect" \cite{S-W}. It is
believed that this is restricted to two dimensions.

The historical remarks about thermal aspects of localization would remain
incomplete without commenting on the \textit{Hawking~radiation}. In that case
the localization results from a reduction of a Hartle-Hawking state on the
global Kruskal extension of a Schwarzschild spacetime to the region outside
the black hole event horizon; the modular Hamiltonian is expected to be
proportional to that of the time-like Killing flow in that region. In the
literature this connection with localized quantum matter is generally not
mentioned; the thermal aspect is often viewed as a thermal manifestation which
can only occur in curved spacetime. Whereas it is certainly true that the
formation of black holes from collapsing stars and the onset of Hawking
radiation is a phenomenon within general relativity, the resulting
thermodynamic equilibrium at the Hawking temperature is a special case of a
thermal manifestation of localization, namely localization of quantum matter
outside a black hole.

The main difference to the B-W situation is that, whereas the Killing time is
the only natural (observer independent) time (corresponding to the Minkowski
time in the zero curvature case), the infinitesimal generator of the
wedge-preserving $z-t$ Lorentz boost is proportional to a Hamiltonian
associated to the proper of a uniformly accelerated observer (radiation
counter) in the z-direction. This is Unruh's realization \cite{Unruh} of
localization within a Rindler-wedge, which, as the result of the impossibility
to accelerate counters to an extend for obtaining a measurable temperature,
remained a "Gedankenexperiment" similar to the E-J conundrum. The connection
of thermal manifestations of quantum matter behind causal and event horizons
with modular theory was given in an important paper of Sewell \cite{Sew}.

The intuitively simple identification of a compact spacetime-localized
observable pictured in the LQP framework as being related to a measurement in
a spatial region with finite duration looses its simplicity; if one tries to
re-express the mathematical description of an ensemble of observables in terms
of physical hardware (radiation counters) in analogy to the Lorentz boost
Hamiltonian of the Unruh effect the intuitive aspect of modular localization
becomes somewhat metaphoric. A concrete realization of a Hamiltonian for
compact localization appears to be impossible. This shows that an intuitively
clear concept, as Haag's localized observables, may take on a metaphoric
appearance if one asks detailed questions about its precise realization in spacetime.

As mentioned before, as long as the mathematical formulation of physical
principles is precise and measurable consequences agree with predictions, such
changes of intuitive arguments under close conceptional scrutiny are of no
concern. QFT is certainly the most successful foundational theory of quantum
matter, but it is also the theory which still contains the largest number of
conceptual surprises. The thermal aspects of the E-J conundrum, which is the
clearest indication that Jordan really discovered a new kind of QT (and not a
relativistic form of QM), is a good illustration.

It has become fashionable to use the word \textit{holistic} for those
properties which set QFT apart from QM in the Fock space formulation
\cite{hol}. The strongest illustration of the holistic aspects of QFT is the
existence of an encoding of a full QFT (including its Poincar\'{e} covariance
acting on Minkowski spacetime as well as its inner symmetries) into an
appropriate relative positioning of a finite number of copies of a rather
structureless monad which act in a shared Hilbert space. Never before in the
history of theoretical physics has a theory been "relational" to such an
extreme degree as in this characterization of QFT. The analogy of such a
presentation with Leibniz's philosophical attempt to understand reality as
resulting from the interplay of impartible objects which he called
\textit{monads} suggests to use this terminology also in the present context
(instead of the rather lengthy name within a classification theory of
indecomposable von Neumann algebras: "hyperfinite type III$_{1}$ factor
algebras"). This is in particular supported by the fact that QFT deals with
only two types: sharply localized monads and fuzzy localized $B(\mathcal{O}%
_{f.l})$ (as well as global algebras $B(H)$). QM (at zero temperature) on the
other hand only uses the $B(H)$ type. The positioning needed in this
construction of a model of QFT is called "modular" (modular inclusions,
modular intersections) \cite{K-W}. The Poincar\'{e} covariance is generated
from the individual modular groups associated with each monad together with
the vacuum state. The inner symmetry is encoded in the superselection sector
determined by the representation classes of the observable net of algebras
\cite{Do-Ro}

Such an encoding of spacetime aspects into a Hilbert space positioning is
impossible in QM; the quantum mechanical localization and the associated
probability is not intrinsic, it has been added to the quantization rules as
an interpretative indispensable tool by Max Born, a step which caused
Einstein's lifelong philosophical dissatisfaction with the post
Bohr-Sommerfeld formulation of QT.

The terminology "holistic" is primarily used in connection with organic
matter. Saying that living things consist of water and certain chemicals is of
little help to understand how they work. In the case of Jordan's fluctuation
model the decomposition into oscillator modes is not wrong but risky because
one is inclined to use approximations which violate the holistic aspect of QFT
without being aware of this; especially if, as in \cite{Du-Ja}, one implicitly
enforces a quantum mechanical setting by stating that the restricted vacuum
remains a pure state. Quantum mechanical computational techniques applied to
infinitely many oscillators are always in danger to violate the holistic
aspects of causal localization. The mathematical raw material (as the shared
Fock space and the momentum space creation/annihilation operators) may be the
same; but important are not the quantum mechanical oscillators themselves, but
rather the resulting holistic object which is obtained with or without their
use. Whereas for free fields and perturbation theory the holistic aspect has
been explicitly worked into the formalism, this cannot be said about the E-J
fluctuation problem.

This has been pointed at in different contexts by other authors. For example
Ehlers (cited in \cite{Du-Ja}) conjectures that Jordan's fluctuation problem
is intimately related to unsolved aspects in the application of QFT to the
problem of the cosmological constant. Hollands and Wald are quite outspoken on
this issue when they say: "Quantum Field Theory Is Not Merely Quantum
Mechanics Applied to Low Energy Effective Degrees of Freedom" \cite{Ho-Wa}.
They use the Casimir effect for their illustration. Additional remarks about
QFT's holistic aspects can be found in the last section.

Jordan may have had a premonition that QFT cannot be subsumed under "QM with
infinite degrees of freedom", but his quantized wave setting was too imprecise
for such distinctions. Knowing these properties with the hindsight of 70 years
of conceptual development in QT, one is inclined to view the critical position
of Jordan's coauthors in a milder light; this was not a fight of a radical
mind against his conservative detractors.

Jordan could not have acquired the status of the \textit{unsung hero of QFT,}
which he earned according to the opinion of historians and philosophers of
physics \cite{Schweber}\cite{Dar}, if he would have merely proposed an
extension of QM to infinite degrees of freedom. In that case he may have
earned the unrestricted support of his collaborators Born and Heisenberg since
the bulk of their joint "Dreim\"{a}nnerarbeit" deals with problems of the
newly discovered QM, but historians may not have considered him the
protagonist of QFT.

The paper is organized as follows. In the next section Heisenberg's area law
of vacuum fluctuations resulting from localized charges is presented in a
contemporary setting of QFT. The third section provides some details on
modular localization of states and operators. In the last section it is shown
that by using the conformal invariance of Jordan's two-dimensional
illustration, in this case the relation to a heat bath thermal system is not
only an analogy but even an isomorphism. The paper concludes with a brief
critical review at developments in which the issue of causal localization was
misunderstood as well as some speculative remarks about future developments of QFT.

In order to keep a check on the length of the bibliography, we refer wherever
this is possible to Haag's book on "Local Quantum Physics". Therefore
citations in the present paper are not directly reflecting the prominence of
their protagonists.

\section{Heisenberg and the localization-caused vacuum polarization}

Of his two coauthors in the Dreim\"{a}nnerarbeit, Heisenberg presented the
strongest resistance against Jordan's claim of the solution of the energy
fluctuation in open subvolume \cite{Du-Ja}. He felt that there were too many
loose ends and uncontrolled assumptions. As time passed, Heisenberg became
more articulate. Beginning in the early 30s he became increasingly aware of a
characteristic phenomenon of QFT, which for composite fields is even present
(although in a milder form) in the absence of interactions: \textit{vacuum
polarization}. Of course the knowledge that the vacuum polarization-caused
impurity of a subvolume-restricted vacuum and its thermal manifestations go
together was not available at that time.

Heisenberg was probably the first who thought that the omnipresence of vacuum
polarization on sharp localization boundaries (endpoints of intervals in
Jordan's simplified chiral current model) create infinities which are only
controllable by making the boundary in some intuitive sense "fuzzy". Nowadays
there is a very precise setting in which this problem allows a rigorous
formulation: the distributional aspect if pointlike fields and the algebraic
\textit{split property}\footnote{It would be interesting to compare the split
property and its resulting area proportionality of entropy to $A/\varepsilon
^{2}$ with $\varepsilon~$the size of the split with 't Hooft's more
speculative brickwall \cite{brickwall} idea which he uses to derive the
Bekenstein area law.}\textit{~}\cite{Haag}. At the beginning vacuum
polarization was understood as a general consequence of systems with infinite
degrees of freedom, but afterwards it was noticed that its appearance is
specific for the causal localization property of QFT; the quantum mechanical
vacuum remains inert even in the limit of infinite particle number
$N\rightarrow\infty$. It may not be obvious to the untrained eye, but the
localization in QM (in second quantization form), where Schr\"{o}dinger
creation and annihilation operators do not appear together in one object (and
where the classical velocity is infinite\footnote{The mean velocity of wave
packets is finite (the acoustic velocity in solids).}), is radically different
from the causal localization of QFT (for which the limiting propagation speed
c is defined algebraically in terms of the causal commutation relations).
Holistic properties do not force an extension of QFT, they only require a
change in our view of QFT.

The simplest illustration of a comparison in terms of shared oscillator-like
momentum space annihilation/creation operators looks as follows%
\begin{align}
a_{QM}(\mathbf{x,}t)  &  =\frac{1}{\left(  2\pi\right)  ^{\frac{3}{2}}}\int
e^{i\mathbf{px-}it\frac{\mathbf{p}^{2}}{2m}}a(\mathbf{p})d^{3}p,~\left[
a(\mathbf{p}),a^{\ast}(\mathbf{p}^{\prime})\right]  =\delta^{3}(\mathbf{p-p}%
^{\prime})\label{field}\\
A_{QFT}(x)  &  =\frac{1}{\left(  2\pi\right)  ^{\frac{3}{2}}}\int\left(
e^{-ipx}a(\mathbf{p})+e^{ipx}a^{\ast}(\mathbf{p})\right)  \frac{d^{3}p}%
{\sqrt{2p_{0}}},~p_{0}=\sqrt{\mathbf{p}^{2}+m^{2}}\nonumber
\end{align}
The global operator algebra generated by smearing with testfunctions of
unrestricted support is in both cases the same, namely the algebra of all
bounded operators $B(H),$ but nothing could be more different than the local
algebras generated with test function of localized support in a compact t=0
region\footnote{As a result of causal relativistic propagation the QFT algebra
is automatically determined in the causal shadow (causal completion)
$\mathcal{O}^{\prime\prime}.$} $\mathcal{C}$. Whereas in the case of QM the
generated operator algebra remains of the same type $B(H(\mathcal{C}))$
(called type I$_{\infty}$ von Neumann factor in a systematic classification of
all von Neumann factor algebras), the local algebras of QFT $\mathcal{A(O})$
defined in terms of the causal completion $\mathcal{O}^{\prime\prime}$ are
factor algebras of hyperfinite type III$_{1}$ which, for reasons explained in
the previous section, will be called \textit{monad}. The only place in the QM
setting where a monad appears is in the thermodynamic limit of thermal Gibbs
systems \cite{Robinson}.

Although the difference in (\ref{field}) appears to be small and consist in a
different Fourier transform of the $a(\mathbf{p})^{\#}$ with a different
energy dependence as well as the appearance of both frequencies in the case of
QFT, and even though the relative equal time commutator effectively limits the
relative nonlocality between the two fields to the size of the Compton wave
length, the consequences of the structural differences are enormous. The
restriction of the vacuum to localized operator algebras of QFT is an impure
KMS state similar to the thermodynamic limit state of QM coupled to a heat
bath, even though in this case there was no coupling to a heat bath. This
property will be presented in more details in the next section.

Another related aspect is the appearance of a vacuum polarization at the
localization boundary of the causal completion (the \textit{causal horizon}).
This phenomenon can also be seen in the behavior of individual operators; they
obey a KMS relation which has no counterpart in QM. Even for free fields this
happens. If one passes to composites as e.g. to a conserved vector currents or
energy-momentum tensor associated with a free field the vacuum polarization
can be seen directly. In this way it was first notice by Heisenberg
\cite{1934}. In fact he correctly guessed already in 1931 in a private
correspondence (see \cite{Du-Ja}) that Jordan may have missed a
log$\varepsilon$ contribution where $\varepsilon$ is the "security distance"
around the end points of the localization interval (which has to be there in
order that the vacuum polarization can "attenuate" and in this way unphysical
ultraviolet divergences can be avoided).

Nowadays such a "fuzzy collar" around a sharp boundary is automatically taken
care of by saying that fields are Schwartz distributions so that smearing
functions which are characteristic functions of a localization regions are
ruled out. This formalism is however not applicable to the localization of
operator algebras, in that case one has to refer to the previously mentioned
split property which also unravel the thermal side of localization.

It is interesting to follow the steps which led Heisenberg to vacuum
polarization in localized operators from a modern\footnote{Here modern means
Schwartz distribution theory, which gave a solid mathematical structure to the
notion of singular quantum fields.} viewpoint. In the classical setting a
conserved charge is the space integral over a conserved current%
\begin{align}
\partial^{\mu}j_{\mu}  &  =0,~Q_{V}^{clas}(t)=\int_{V}d^{3}x~j_{0}%
^{clas}(t,\mathbf{x})\\
Q_{V}^{QM}(t)  &  =\int_{V}d^{3}x~j_{0}^{QM}(t,\mathbf{x}),~Q_{V}%
^{QM}(t)\Omega^{QM}=0\nonumber
\end{align}
The partial charge in the volume $V$ is still t-dependent and becomes a
dimensionless time-independent constant in the limit $V\rightarrow\infty.$ The
partial charge in second quantized charged Schroedinger QM can be defined in
the same way apart from the fact that $Q_{V}(t)$ is now an \textit{operator}
which annihilates the quantum mechanical vacuum $\Omega^{QM}$. The situation
changes radically in QFT where this way of writing does not make sense because
quantum fields are by there very nature rather singular object (they are
operator-valued distributions). The degree of singularity follows in this case
from the property that a charge must be dimensionless $(d(Q)=0)$ and hence
$d(j_{\mu})=3$. In scale-invariant theories this would fix the inverse short
distance power in $x$, whereas in massive theories it determines only the
short distance singular behavior of conserved currents.

Heisenberg's important observation which led to the terminology "vacuum
polarization" was made formally, i.e. without the use of test functions
(distribution theory came two decades later) on "partial" charges associated
to conserved currents of charged (complex) relativistic free fields (s=0,1/2).
The conserved current is a charge neutral bilinear local composite which,
apart from the Wick-ordering of the involved product of free fields, is equal
to the classical Noether expression, and its application to the vacuum creates
neutral pairs of particles (hence the term "vacuum polarization"); in the
presence of interactions their number is always unlimited ("polarization
clouds"). \ 

Heisenberg found that if one integrates the zero component of the conserved
current of a charged free field over a finite spatial region of radius R, the
so defined "partial charge" diverges as a result of vacuum polarization at the
boundary, and that only by integrating all the way to infinity one obtains a
finite polarization-free global charge of the respective state on which the
operator is applied.

In the modern QFT setting it is possible to control the strength of such a
divergence in terms of specially prepared test functions. Elementary and
composite fields are singular but very precisely defined objects, they are
operator-valued distributions. Whereas in QM the knowledge of the use of the
Dirac delta functions suffices, the correlation functions of fields in QFT and
their handlings necessitate at least a rudimentary knowledge of Schwartz
distribution theory which exists since the beginning of the 50s and played an
important role in Wightman's approach to QFT. A finite partial charge in $n$
spacetime dimensions with a vacuum polarization cloud within a spherical
region of thickness $\Delta R$ is is defined in terms of the following
\textit{dimensionless} operator%
\begin{align}
&  Q(f_{R,\Delta R},g_{T})=\int j_{0}(\mathbf{x},t)f_{R,\Delta R}%
(\mathbf{x})g_{T}(t)d\mathbf{x}dt,~lim_{R\rightarrow\infty}Q(f_{R,\Delta
R},g_{T})=Q\label{partial}\\
&  \left\Vert Q(f_{R,\Delta R},g_{T})\Omega\right\Vert \equiv F(R,\Delta
R)\overset{\Delta R\rightarrow0}{\sim}%
\genfrac{\{}{.}{0pt}{}{C_{n}(\frac{R}{\Delta R})^{n-2}\text{ }%
for~n>2}{C~ln(\frac{R}{\Delta R})~for~n=2}%
\nonumber
\end{align}
where the spatial smearing is in terms of a test function $f_{R,\Delta
R}(\mathbf{x})$ which is equal to one inside a sphere of radius R and zero
outside $R+\Delta R,$ with a smooth transition in between; and $g_{T}$ is a
finite support $\left[  -T,T\right]  $ interpolation of the delta function. As
a result of current conservation such expressions converge\footnote{However
inside commutators with other localized operators the partial charge is
already time-independent as soon as their causally completed localization
region is contained in that if the partial charge. In this algebraic sense the
partial charge is really "partially" time-independent.} with this special
choice of "smearing" functions for $R\rightarrow\infty$ independent of
$\varepsilon$ to the global charge either weakly \cite{Sw} or (of one sends T
together with R appropriately to infinity) even strongly on a dense set of
states \cite{Requ}. This aspect of the quantum Noether issue is quite
important because in addition to the "normal" behavior there are two other
cases which have no classical or quantum mechanical counterpart.

The first case which has been exemplified by Goldstone leads to a divergence
of the partial charge in the limit of $R\rightarrow\infty$ due to a zero mass
Goldstone particle \cite{Haag} which couples to the conserved current and
prevents the convergence to a finite limit and is referred to as "spontaneous
symmetry breaking". The second case is the Schwinger-Higgs "charge screening"
in which the sequence of partial charges $Q(f_{R,\Delta R},g_{T})$ converges
to zero; in this case there is no (charge) symmetry to start with which can be
broken. This is important for describe massive vectormesons in the setting of
renormalized nonabelian gauge theory \cite{integrable}. Often the
Schwinger-Higgs mechanism is erroneously referred to as a symmetry breaking
(which symmetry? gauge transformations are not physical symmetry
transformations the transition from charge conservation to zero charge can
hardly be called a symmetry breaking). A theorem relates the transition from
massless to the massive vectormeson with the charge screening \cite{Sw}%
\cite{nonloc}.

For an estimate of the vacuum polarization one is interested in the limit of
$\Delta R\rightarrow0$ for fixed $R$ of $F(R,\Delta R)$ (\ref{partial}). As
expected and already argued by Heisenberg, the computation of the partial
charge in Jordan's chiral conformal model shows a logarithmically divergent
behavior, whereas for n-dimensional models in case of n%
$>$%
2 the vacuum fluctuations in the fuzzy boundary are proportional to the
"dimensionless area" $\frac{area}{\left(  \Delta R\right)  ^{n-2}}$ which
diverges in the limit of sharp localization $\Delta R\rightarrow0$ of the
partial charge as in (\ref{partial}). The calculation is particular simple in
the massless conformal limit of a conserved current. The same limiting
behavior which appears in the dimensionless partial charge also shows up as
the leading short distance terms in the (also dimensionless)
\textit{localization entropy} \cite{BMS} which refers to a fuzzy localized
operator algebra instead of a single operator (next section).

The correct treatment of the perturbative vacuum polarization contributions
was a painful process which almost led to the abandonment of QFT (the
ultraviolet crisis of QFT). The importance of causal localization as the
holistic principle which separates QFT from (infinite degree of freedom) QM
and has to be upheld even in approximations was certainly not known at that
time. Eventually a formulation was found which avoids intermediate violations
of locality and Hilbert space requirements (positivity) caused by the cutoffs
and regulators of the old renormalization method and instead presents
renormalized perturbation theory directly in terms of its foundational root as
an iterative implementation of the causal localization principle together with
the requirement of a maximal scaling degree\footnote{The bound on the scaling
degree can only be fulfilled if the lowest order interaction in terms of local
Wickproducts of free fields has scaling degree $\leq4.$ The resulting finite
parametric expression defines an island in the infinite parametric result
(obtained without the scaling restriction) which is left invariant under
renormalization group transformations.}. As an implementation of a principle,
this Epstein-Glaser approach \cite{E-G} is free of any intermediate
ultraviolet divergences resulting from a hidden incorrect handling of vacuum
polarization which plagued older formulations.

It is believed that among all interacting models which allow a
characterization in terms of Lagrangian presentation, only the renormalizable
models have the chance to be supported by a future mathematical existence
proof. But since the perturbative series diverge, this has remained one of the
great unsolved problems which has no counterpart in any other area of
theoretical physics (where it was always possible to find mathematically
controlled approximations). In this connection Gedankenexperiments as the
Einstein-Jordan conundrum (or Unruh's localization in the Rindler wedge in
terms of accelerated observers) are valuable because they point towards
another face of QFT which, even after almost a century, simply remained
outside the range of the standard quantization formalism. Hence such
Gedankenexperiments may reveal good reasons why our conceptual mathematical
access remained insufficient for establishing the mathematical existence of
quantum models behind classical Lagrangians.

The heuristic aspects of Lagrangian quantization and functional integral
representation are sufficient for starting renormalized perturbation theory,
but they do neither help in conceptual mathematical problems of establishing
existence nor are they even useful to understand thermal and entropic aspects
of localization. Problems as the Einstein-Jordan conundrum remind us of
unfinished business which holds QFT back from its closure.

The standard model of a QFT is one with a complete particle interpretation
i.e. one in which fields are related via large time scattering asymptotes to
particles and the full Hilbert space is a Wigner Fock space generated by those
in/out particles. In such models the fact that particles are directly related
to measurable quantities as scatting amplitudes and the closely related
formfactors is of considerable practical and theoretical value. An intuitive
argument, which relates properties of formfactors which are the n-particle
components of a state $A\left\vert 0\right\rangle $ obtained by
"banging"\footnote{Causally localized operators applied to the vacuum
("bangs") create states with the full energy spectrum; in the presence of
interactions these states also contain the full in (or out) particle
spectrum.} with a local operator $A\in\mathcal{A(O})$ on the vacuum
$\left\vert 0\right\rangle $, is based on the following relation%
\begin{align}
&  A\left\vert 0\right\rangle \rightarrow\left\{  \left\langle p_{1}%
,p_{2}...p_{n}\left\vert A\right\vert 0\right\rangle \right\}  _{n\in
\mathbb{N}}\label{cross}\\
&  \left\langle 0\left\vert A^{\ast}\right\vert p_{1}p_{2}...p_{n}%
\right\rangle ^{in}=~^{out}\left\langle -p_{n},-p_{n-1},...-p_{k+1}\left\vert
A^{\ast}\right\vert p_{1}p_{2}...p_{k}\right\rangle ^{in}\nonumber
\end{align}
\newline where the second line is the \textit{crossing identity} (last
section). In other words, a "bang" on the vacuum leads to a state with an
arbitrary high number of particles, or to phrase it in the vernacular manner
of Murphy's law: \textit{what can occur} (as the outcome of a bang subject to
the superselection rule) \textit{actually does occur}. To act on the vacuum
more softly, so that (as in the case of partial charges) the excitation of
states with arbitrary high particle number is suppressed, one has to resort to
"quasilocal operators" \cite{Haag}. The $-p$ refers to a well defined analytic
continuation from the positive mass shell to the negative mass shell so that
the last aspect of virtuality of (\ref{cross}) is removed by stating that the
actual particle production in a bang on the vacuum is uniquely determined by
the kind of "hammer" $B$ used for banging\footnote{Depending on its more or
less sharp surface (fuzzy boundary) the vacuum polarization clouds are
stronger of weaker.}; or returning to more scientific parlance, the vacuum
formfactor of a local operator B determines its associated general formfactor
(with the same total number of particles) by analytic continuation.

The "bang on the vacuum" concept is less metaphoric than the picture of the
QFT vacuum as a "broiling soup" of virtual pairs \cite{Summers}. No conceptual
headstand like "allowing the uncertainty relation for a short time to
invalidate the energy conservation" is necessary. In fact the uncertainty
relation is connected with the position operator which is not a well-defined
(frame-independent) object in QFT; hence an uncertainty relation has no
conceptual place in QFT since there is no position operator. Its covariant QFT
counterpart is the increase of the localization entropy/energy with the
sharpening of localization by compressing the fuzzy surface sheet of size
$\varepsilon$ in definitions of partial charges (\ref{st}) and
localization-entropy (see the next section).

It is quite interesting to add another fact about the local banging. Whereas
the application of all global operators generates the full Hilbert space, the
application of a local algebra $\mathcal{A(O})$ does not, as one may naively
think, generate a closed subspace as it does in QM; but rather generates a
dense subspace of $H$ (the Reeh-Schlieder theorem \cite{Haag}) which changes
with the localization region without loosing its property of being dense. This
confirms that by "changing the hammer" within $\mathcal{O}$ one gets
arbitrarily close to any particle state. Although this does not yet reveal the
thermal aspects of localization, it does point to another property which shows
that behind Haag's visualizing of local observable in terms of finite
extension and duration of measurements loom quite metaphoric details if one
insists to re-express its mathematical precision in terms of idealized
manipulations on experimental hardware. The only understood case is that of a
localization in a wedge (Unruh's accelerated observers). But as previously
stated, metaphoric aspects of foundational concepts are perfectly acceptable
as long as the mathematical consequences can be clearly formulated. In fact
from a philosophical viewpoint one would even expect that this discrepancy
between the intuitive content of principles and the precise reformulation of
their mathematically rigorous setting increases as theories become more
inclusive and fundamental. For the case of compact localization this process
of reformulation in terms eludes a visualization in terms of physical hardware.

The great remaining problem which will decide the future of QFT is therefore
to find nonperturbative techniques which are in accordance with the intrinsic
holistic localization of QFT and are mathematically controlled. For this we
first have to understand in more detail what this intrinsic nature of QFT
consist of. The finding will enable us to decode the E-J conundrum (section 4).

\section{Local quantum physics: modular localization and its thermal
manifestation}

Although by 1929 \cite{Kharkov} Jordan knew through previous work with Pauli
that QFT led to manifestations which were distinctively different from those
of infinite degree of freedom QM, the idea that one needs a new conceptual
setting did not yet take hold. From the viewpoint of the formalism of
Lagrangian quantization there was no visible distinction, except that
relativistic QFT was Poincar\'{e} covariant and, at least in the presence of
interactions, did not allow a "first quantized" wave function formulation.
Apart from occasional flare-ups which found their expression in sayings like:
"putting QFT on its own feet", or " QT without classical (quantization)
crutches" \cite{Kharkov}, there was as yet no concerted effort to understand
both quantum theories in terms of different intrinsic principles rather than
of a shared quantization formalism. The conceptual-mathematical setting of QM
reached its closure already in the 30s in the work of John von Neumann and
Hermann Weyl before it also entered textbooks. At that time the not even the
vacuum polarization phenomenon of QFT was properly understood.

Foundational work on QFT started more than two decades later, partly because
the fundamental differences were initially not perceived as such. Without this
move the old ultraviolet infinities, which plagued QFT for more than a decade,
would have continued to be a cause confusion. The number of renormalizable
couplings of pointlike fields in d=1+3 is finite and all of them are known.
The standard model, i.e. the joining of weak, electromagnetic and string
interactions under the roof of gauge theories, was the last big push; after
this, particle theory entered an already 40 years lasting period of conceptual stagnation.

The only idea which requires a different (not yet completely elaborated) form
of perturbation theory is the use of string-localized fields of higher spin
\cite{MSY}\cite{Yng} which formally improve their short distance behavior and
enlarges the number of possible models which fulfill the power counting
renormalizabilty criterion. For the s=1 gauge theories their subset of
point-localized fields agrees with the gauge invariant fields of the
quantization gauge approach. The use of string-localized potentials leads to
fields in positive metric Fock-spaces with short distance dimension
$d_{s.d}=1$; this enlarges the potentially renormalizable couplings to
infinitely many possibilities. Details about how modular localization leads to
these new string-localized fields and what can be expected from their
perturbative use can be found in \cite{char} \cite{nonloc} \cite{Yng}
\cite{integrable}.

In Haag's approach a QFT is defined in terms of a net of local (von Neumann)
operator algebras $\left\{  \mathcal{A(O})\right\}  _{\mathcal{O}\subset
M_{4}}.$ $~$Fields in the sense of Wightman are global objects $\Phi(x)$ which
upon smearing with $\mathcal{O}$-supported test functions $suppf\subset
\mathcal{O}$ become localized unbounded operators $\Phi(f)$ affiliated to
local algebras $\mathcal{A(O}).$ The global nature of the field is reflected
in the fact that it serves as a \textit{generator for all local algebras}. One
believes that all physically relevant nets of local algebras are generated by
local fields, but the lack of a general proof is not very important because
the algebraic setting has been shown to contain the full interpretation of the
theory, in particular the important relation to particles in terms of
scattering theory \cite{Haag}.

The algebras fulfill a set of obvious consistency properties which result from
the action of Poincar\'{e} transformations, spacelike commutation relations
and causal completeness properties. Instead of the Poincar\'{e} transformation
law of covariant spinorial $\Psi^{(A,\dot{B})}$ fields, it is only required
that the transformed operator $A\in\mathcal{A(O)}$ belongs to the operator
algebra of the transformed region. The causality requirements are%
\begin{align}
\left[  A,B\right]   &  =0,~A\in\mathcal{A(O)},~B\in\mathcal{A(O}^{\prime
}\mathcal{)}\subseteq\mathcal{A(O})^{\prime},~Einstein~causality\label{sha}\\
\mathcal{A(O})  &  =\mathcal{A(O}^{\prime\prime}%
),~causal~shadow~property,\text{ }causal~completion~of~\mathcal{O}\nonumber
\end{align}
Here the first line the algebraic formulation of the statistical independence
of spacelike separated events; the upper dash on the spacetime region denotes
the spacelike disjoint region, whereas on the algebra it stands for the
commutant algebra. The second line is the local version of the aforementioned
time-slice property \cite{H-S} where the double causal disjoint $\mathcal{O}%
^{\prime\prime}$ is the causal completion (shadow) of $\mathcal{O}.$

One of the oldest observations about peculiar consequences of causal
localization in QFT is the Reeh-Schlieder property \cite{Haag} i.e. the
denseness of $\mathcal{A(O)}\left\vert 0\right\rangle $ in $H.$ Together with
Einstein causality this denseness property leads immediately to the
\textit{standardness} of the pair ($\mathcal{A(O}),\left\vert 0\right\rangle
$) i.e. the property that $\mathcal{A}(\mathcal{O})$ acts cyclic (the density
property) and (in contrast to the global algebra) separating (contains no
annihilators) on $\left\vert 0\right\rangle .$

This property attracted the attention of philosophical inclined particle
physicists as no other quantum aspect. Its counterintuitive aspect, admitting
to change the situation "behind the moon" by doing something in an arbitrary
small laboratory during an arbitrary small duration, is in a way more radical
than even the quantum mechanical "Schroedinger cat" paradox and the EPR
entanglement\footnote{In quantum information theory the word "nonlocal" has a
different meaning than in QFT.} phenomenon.

As mentioned before, in ("the second quantized" formulation of) QM
localization is related to the spectral theory of a position operator and
refers to the subspaces associated with the projectors of its spectral
decomposition; the objects which are being localized in a spatial region
$\mathcal{O}$ are subspaces $H(\mathcal{O})\subset H$ or state vectors
$\psi(f)\left\vert 0\right\rangle $ in which the test functions have their
support in the spectrum of the selfadjoint position operator $spec\mathbf{x}%
\subset R^{3}.~$This state-localization, which is directly related to Born's
probability concept, plays also a role for relativistic wave function. As
Newton and Wigner showed \cite{Haag}, it depends on the frame of reference
i.e. is not part of the Lorentz covariant observables of QFT. In the setting
of "direct particle interaction" (DPI) \cite{Coe} \cite{interface}, a kind of
relativistic QM with interaction, the covariance is only recovered in the
scattering limit i.e. the resulting M\"{o}ller operator and its S-matrix are
the only covariant operators, in fact they are even invariant as they should
be. Whereas DPI is primarily a quantum mechanical construction of a
relativistic S-matrix, QFT is a theory of covariant localized observables and
the invariance of the S-matrix is a direct consequence of the covariance of
local observables. In QFT there are covariant fields but no particles at
finite times; particles and fields harmonize only asymptotically; any attempt
to enforce their coexistence at finite times in the presence of interactions
is bound to fail (see below). In DPI particles exist at all times at the prize
of absence of covariant observables besides the global S-matrix.

The Born localization is also important for QFT where its range of application
is limited to wave functions. It turns out that the centers of wave packets
which mark the region of largest Born probability density for large times
follow linear orbits with sharply defined velocities which agree with what one
expects from the result from sharp causal propagation i.e. in this asymptotic
(effective) sense both localizations coalesce. All the alleged superluminal
violations of causality which appeared and still appear in articles and
journals have their origin in the incorrect identifications of the two
localization concepts for finite times. The velocity of sound in QM or the
limiting velocity $c$ in DPI only attain their precise meaning (sharp value)
at large times.

The mathematical backup of the QFT localization is the \textit{modular
operator theory} in the setting of operator algebras (for references consult
\cite{interface}). Its most important operator is the unbounded involutive
Tomita operator $S_{\mathcal{O}}$ (see below)~whose domain $domS_{\mathcal{O}%
}~$is intimately related to the Reeh-Schlieder dense subspaces. A closer
examination of the Tomita $S$-operators (next section) in case of
$\mathcal{O}=W$ reveals that their domain is entirely determined by the
representation of the Poincar\'{e} group, a property which is shared between
the interacting and their associated free incoming operators. The interaction
only enters in the specific way $S_{W}$ maps vectors from $domS_{W}$ into
their image in $S_{W}$.

As already mentioned, the conceptual consequences of this theorem exerted an
enormous attraction to philosophers of science as can be seen from articles
under the heading "Reeh-Schlieder wins against Newton-Wigner" \cite{Hal} which
if phrased in the terminology of the present article would read "modular
localization versus Newton-Wigner localization". This antagonistic title
represents the "half glass empty view" of the situation; in the "half glass
full view" of the present paper one would instead emphasize that the two
localizations become compatible at the only place where it really matters,
namely in the asymptotic scattering region \cite{interface}.

In the presence of modular localization it is less important to understand the
individual differences of operators within a local algebra, since the fact
that they share the localization in a spacetime region $\mathcal{O}$ and that
the state vectors they create from the vacuum have a nonvanishing inner
product with all n-particle states already suffices to extract the S-matrix
\cite{Haag}. LQP derives all properties of particle theory from ensembles of
operators and (apart from generating conserved currents which generate
symmetries) sidesteps properties of individual operators. Higher precision in
characterizing observables belonging to specific spacetime regions amount to
improvement in localization, just as in real experiments where the precision
in measuring localizations of charges, momenta, masses and spins depend on the
improvement of the internal structure of counters and the geometric relation
between sources and counters. As explained in the first section, the
appropriate framework for formulating causality properties is that of nets of
spacetime-indexed \textit{von Neumann algebras}; only in such operator
algebras can the weak closure referring to states be replaced by the double
commutant operation i.e. $\mathcal{A}=\mathcal{A}^{\prime\prime}$ (which is a
pure algebraic concept) and the causal completion matches the algebraic notion
of commutant used in the formulation of Einstein causality. These operator
algebras may be obtained from the more abstract C$^{\ast}$-algebras by
representation theory, a viewpoint which is particularly helpful in the
presence of unitarily inequivalent representations. As previously mentioned
the modular aspects of the operator algebraic setting entered particle physics
for the first time in the conceptually correct formulation of the problem of
\textit{statistical mechanics of open systems~}\cite{HHW}.

Already at the beginning of the 60 it was known that local algebras could not
be of quantum mechanical type I$_{\infty}$. At the time of the first LQP paper
\cite{H-S} after Haag's talk at the 1957 Lille conference \cite{Lille}, most
people (including myself) still tacitly believed that the type should be
I$_{\infty};$ in part because almost nothing was known about type III algebras
and the only kind of algebra one met in QM was that of all bounded operators
$B(H)$ in a Hilbert space which is I$_{\infty}$. Two years later Araki showed
that it belongs to the type III family \cite{Haag}. A decade later, after the
refinement of the type classification by Connes, an important addition by
Haagerup and the geometric physical identification of the modular objects of
wedge-localized QFT operator algebras $\mathcal{A}(W)$ by Bisognano and
Wichmann \cite{Bi-Wi}, all the concepts which link localization in QFT with
thermal aspects and made the monad unique (it represents just one isomorphy
class) were in place.

Around the same time there were independent observations about QFT in curved
space time (CST) \cite{Haag}, more precisely restrictions to partial spacetime
regions in front of black hole event horizons. It was noted that, unlike
classical matter, the presence of quantum matter leads to Hawking radiation at
the Hawking temperature. A similar thermal manifestation was pointed out in
the setting of a Gedankenexperiment by Unruh \cite{Unruh}. In order to
localize an observer in such a way that his accessible spacetime region is a
Rindler wedge W, he has to be uniformly accelerated in which case his world
line is traced out by applying the wedge-related Lorentz boost to his start at
t=0 inside the wedge. His proper time is different from the Minkowski time and
depends on the constant acceleration. Taking this into consideration, the
observer's Hamiltonian is only different from the dimensionless generator of
the Lorentz boost by a dimensionfull numerical factor.

Even if the tiny associated temperature will never be measured and Unruh's
proposal always remain a Gedankenexperiment, the consequences of modular
localization for QFT are of pivotal structural importance for QFT. As will be
seen they are not only helpful in order to completely resolve the
Einstein-Jordan conundrum at the cradle of QFT, but even more important to
understand the conceptual basis of subtle properties of particle theory as the
particle crossing property of formfactors and the existence of nontrivial
models and controlled ways to approximate them. At present there is no other
setting than that of modular localization which has a chance to place QFT side
by side to all the other already conceptually closed theories in the pantheon
of theoretical physics.

The Unruh Gedankenexperiment is a special case of a more general setting of
modular localization\footnote{This was pointed out by Sewell \cite{Sew} (but
apparently not accepted by Unruh who maintained that the modular theory has
nothing to do with the effect which bears his name).} \cite{S2} which
describes the position of the dense subspace in terms of domains of the
unbounded Tomita involution $S$. These domains are determined in terms of the
unitary representation of the Poincar\'{e} group, but for knowing the action
of the operator itself, one needs to know dynamic aspects of a QFT which leads
to those localized states. This localization is intrinsic to the Wigner
representation theory for positive energy representation. In the absence of
interactions there is a direct functorial passage from subspaces of states to
subalgebras. Since this has been explained in detail in \cite{BGL}%
\cite{Fa-Sc}\cite{MSY}\cite{nonloc} and since the operator algebra
localization for interacting QFT is more important in the present context than
that of subspaces, we refer the reader to the literature.

The general modular operator theory starts with the definition of the Tomita
$S$-operator (section 1, \ref{S} \ref{T-T}) and the action of the operators
$\Delta^{it},J$ on the algebra $\mathcal{A}~$obtained from the polar
decomposition of $S.$ The prerequisite of "standardness" for the applicability
of the T-T modular theory to localized algebras of QFT are always fulfilled
thanks to the Reeh-Schlieder theorem \cite{Haag}. It guaranties the universal
validity of modular theory for local algebras of QFT with respect to any
finite energy state as long as their causal closure is not the global algebra
(which contains all operators and hence in particular annihilators of finite
energy states). Unless specified otherwise, $\Omega~$in the sequel denotes the
vacuum. In fact the $domS$ is nothing else than the closure of the
Reeh-Schlieder subspace in the graph norm of $S.$

In QFT the $\Delta^{it}$ for the case of $\mathcal{O}=W$ is kinematic in the
sense that it only depends on the representation of the Poincar\'{e} group
$\Delta_{W}^{it}=U(\Lambda_{W}(-2\pi t))$. Hence it is shared between all QFTs
which live in the same Hilbert space and in addition of sharing the vacuum
have the same particle content. On the other hand the modular reflection $J$
and therefore $S$ depend on the interaction through the $S_{scat}$ matrix
\begin{equation}
J_{W}=J_{0,W}S_{scat} \label{scat}%
\end{equation}
where the subscript $0~$\ refers to the interaction-free algebra generated by
the incoming fields and $S_{scat}~$is the scattering matrix. This follows from
the work of Jost \cite{Jost} on the TCP operation; is constructive use in for
model constructions was noted in \cite{S2}. In this case of $\mathcal{O=}$ $W$
the $\Delta^{i\tau}$ is (up to a scaling factor) the W-preserving Lorentz
boost. In fact according to the B-W theorem \cite{Haag} $\Delta_{W}^{i\tau
}=e^{-i2\pi\tau K}~$with~$K=$ infinitesimal generator of the W-preserving
Lorentz boost, and $J=TCP$ up to a $\pi$-rotation around the z-axis.

The central result of the Tomita-Takesaki modular theory is the T-T theorem
which states that $\Delta^{i\tau}$ defines a (modular) automorphism of the
operator algebra and $J$ an anti-isomorphism into its commutant%
\begin{equation}
\Delta^{i\tau}\mathcal{A}\Delta^{-i\tau}=\mathcal{A},~J\mathcal{A}%
J=\mathcal{A}^{\prime}%
\end{equation}

Although the domain of the Tomita $S$-operator (except for wedge-localized
algebras), allows no direct description in terms of the Poincar\'{e} group for
subwedge regions, this information can be obtained from intersections of the
real subspaces $\mathcal{K}$ associated with $S_{W}$%
\begin{equation}
\mathcal{K}_{W}=\left\{  \psi;S_{W}\psi=\psi_{W}\right\}  ,~\mathcal{K}%
_{\mathcal{O}}=\cap_{W\supset\mathcal{O}}\mathcal{K}_{W}%
\end{equation}
and hence even these domains are (indirectly) of kinematic origin. The dynamic
content of subwedge reflections $J_{\mathcal{O}}$ is however not known.

The general modular localization situation is more abstract than its
illustration in the context of the Unruh Gedankenexperiment since the generic
modular Hamiltonian is not associated with any spacetime diffeomorphism; it
rather describes a "fuzzy" movement which respects the causal boundaries (the
horizon) but acts somewhat nonlocal on the inside bulk; very little is known
about properties of modular Hamiltonians. But even in case where the analog of
a localized Unruh observer is not available, the mere knowledge about the
\textit{existence} of a modular Hamiltonian is of great structural value,
since it allows to give a mathematical precise quantum physical description of
the locally restricted vacuum as an impure (singular KMS) state associated
with the intrinsically determined modular Hamiltonian. The KMS property
formulated in terms of the restriction of the vacuum to the algebra of the
localization region reads%
\begin{align}
\left\langle AB\right\rangle  &  =\left\langle Be^{-H_{mod}}A\right\rangle
,~\ \Delta=e^{-H_{mod}},~A,B\in\emph{A(}\mathcal{O})\label{KMS}\\
\left\langle AB\right\rangle  &  \neq\left\langle A\right\rangle \left\langle
B\right\rangle ~~if~\left[  A,B\right]  =0~in~contrast~to~QM\nonumber
\end{align}

Here the (unbounded) modular operator $\Delta$ and its associated modular
Hamiltonian $H_{mod}$ are associated to the "standard pair" $(\mathcal{A(O)}%
,\Omega)$. In the case of heat bath thermality (statistical mechanic) the
Hamiltonian is the standard Hamiltonian which implements time translations in
Minkowski spacetime, $\mathcal{O}$ is replaced by the global spacetime
$\mathbb{R}^{4}$ and the vacuum state $\Omega$ is now the GNS vector
\cite{Robinson} of the thermal equilibrium state. The KMS property associated
with the thermal manifestation of the vacuum setting of QFT serves primarily
to direct attention to impure KMS nature of restricted vacua. The modular
Hamiltonian changes if one enlarges the algebra by adding observables
localized outside.

Since two localized operators $A,B\in\mathcal{A(O})$ belong to a continuous
set of algebras with larger localization$,$ there is a continuous set of
modular Hamiltonians which lead to KMS commutation relations with different
$\Delta$ for the same localized pair of operators. This leads to a continuous
infinity of KMS relation for a given pair and shows that QFT is a much tighter
and more fundamental theory than QM. This rather subtle property of QFT
illustrates again that, despite the immediate intuitive appeal of Haag's LQP
setting, its mathematical consequences are anything but intuitive.

The thermal KMS property as a consequence of modular localization is primarily
an attribute of an ensemble of observables which share the same localization
region. As a consequence it is also a relation which each individual operator
obeys, but the probability notion coming with thermal ensembles always reminds
us that, unlike the situation in QM, the ensemble interpretation is intrinsic
and does not have to be added. The important point here is that it would have
been difficult to discover this infinite set of relations without knowing
anything about the nature of modular localization within Haag's setting of LQP
\cite{Haag}.

It is not conceivable that with the full knowledge about the thermal
consequences of quantum causal localization Einstein would not have accepted
this probability since he used probabilities of thermal ensembles in his
fluctuation arguments for the existence of photons. QM is a global setting and
its Born-localization and related probability based on the spectral
decomposition of a position operator is ostensibly referring individual
events. If the less fundamental QM could be derived from QFT in a limit which
maintains the QFT probability but destroys its modular localization including
its concomitant thermal manifestation one may even speculate that Einstein may
have reconciled himself with Born's probabilistic interpretation of events in QM.

The old pre-renormalization perturbation theory failed precisely because
quantum mechanical perturbation methods were used which do not keep proper
track of covariance and causality. The covariant formulation at the end of the
40s consisted of recipes in which the role of the causal locality principle
was not clearly visible so that problems as subvolume energy fluctuations (for
which covariance was not of much help), lacked a precise conceptual
understanding. In the case of the oscillator decomposition of Jordan's quantum
field theoretic fluctuation model, the approximation of only occupying
oscillator levels in a certain frequency range generally destroys the
localization and thermal aspects required by QFT \cite{Du-Ja}. In an unguided
oscillator approximation of the subvolume fluctuations one almost certainly
destroys the holistic localization. The way to maintain it is the
implementation of the LQP split property \cite{Haag}; but since this refers to
operator algebras, this leads to problems which are more difficult than the
distribution-theoretical single operator calculation of "fuzzy" boundaries in
the calculation of partial charges in section 2.

It was noted elsewhere \cite{Ho-Wa} that the holistic aspect of modular
localization renders the use of quantum mechanical ideas of global level
occupation (used e.g. in some estimates of cosmological vacuum energy density)
potentially misleading\footnote{It is interesting that Ehlers (see
\cite{Du-Ja}) mentions the problem of cosmological constant in connection with
the Einstein-Jordan conundrum, thus suggesting that in both problems the role
of the vacuum polarization has not been properly understood.}.

Among the properties which cannot be ascribed to an individual operator but
corresponds (similar to the E-J fluctuation) to a localized algebras is the
problem of \textit{localization-entropy} as a measure of localization-caused
vacuum polarization. The entropy of sharp localization is infinite; unlike the
infinities in the old ultraviolet catastrophe this is a genuine infinity in
QFT; in fact in the absence of a position operator it represents the QFT
analog of the QM uncertainty relation. There is a close connection with the
infinite volume infinity of the heat bath entropy. For chiral theories on the
lightray there is a rigorous derivation of the well-known linear increase (the
"one-dimensional volume factor" $L$) of the heat bath entropy and the
logarithmic growth of the QFT entropy with decreasing attenuation distance
$\varepsilon$ of vacuum polarization which are related\footnote{In a more
detailed description of chiral theories the fuzzyness $\varepsilon~$can be
expressed in terms of a conformal invariant ratio of 4 points which are the
end points of a smaller interval included in a bigger one \cite{BMS}.} by
-ln$\varepsilon\sim L~$as a result of a conformal isomorphism (next
section)$.$

In higher dimensions there are rather convincing arguments that the limiting
behavior for $\varepsilon\rightarrow0$ for the dimensionless entropy is the
same as in the increase of the dimensionless partial charge (\ref{partial}).
For the same geometric situation as in case of the partial charge
(\ref{partial}) this suggests%

\begin{equation}
V_{n-1}\left(  kT\right)  ^{n-1}|_{T=T_{\operatorname{mod}}}\simeq%
\genfrac{\{}{.}{0pt}{}{ln\left(  \varepsilon^{-1}\right)  ,~n=2}{\left(
\frac{R}{\Delta R}\right)  ^{n-2},~n>2}
\label{st}%
\end{equation}
where the volume proportional (dimensionless) entropy on the left hand side is
the standard heat bath entropy; the n-2 power on the right hand side represent
the n-2 transverse directions in n dimensions. Whereas, as a result of the
existence of an inverse Unruh effect \cite{S-W}, the derivation of the n=2
localization is a consequence of that isomorphism, the n%
$>$%
2 case is supported by the analogy to the partial charge. It also agrees with
't Hooft area behavior from the brickwall assumption \cite{brickwall} which is
matched with Bekenstein's classical area law by using for the free parameter
$\varepsilon$ the numerical value of the Planck length.

Since a rigorous implementation of the split formalism is still missing, there
is room for another idea which amounts to a multiplicative logarithmic
modification for n%
$>$%
2 by $ln\frac{R}{\Delta R}.~$In that case the box-volume of the heat bath side
would correspond to a volume of a box for which two sides are transverse and
one (which accounts for the $ln$ factor) would be lightlike, as in the chiral
case n=2. This could be seen as a weak version of an \textit{inverse} Unruh
situation: there is no isomorphism between the two systems but there still
remains a close analogy between heat bath and localization-caused thermal
behavior \cite{BMS}. From a mathematical viewpoint this suggests that the
monad in its role of describing a thermodynamic limit of heat bath system and
that used for localization are mainly different in terms of their different
physical parametrization.

The strongest illustration of the holistic aspects of QFT as compared to QM is
the characterization of models of QFT (including the quantum matter content as
well as its ordering spacetime symmetry structure) in terms of modular
positioning which was already mentioned in the introduction. This was first
observed in chiral QFT on the lightray, permit a characterization in which
instead of spacetime (interval) ordered quantum matter (nets of operator
algebras) permitted a complete characterization in terms of the relative
positioning of a finite number of monads. After preliminary observation on
"quarter circle inclusions" \cite{quarter}, the appropriate mathematical
setting was found in terms of the notion of \textit{halfsided modular
inclusions \cite{Wies}, i.e. }inclusions of one monad in another which share
their standard vector and for which the modular group $\Delta^{it}$ of the
bigger monad compresses the smaller one for one sign of t. It turns out that
the monad requirement on the operator algebras can be omitted; it follows from
the halfsided modular property of the inclusions; the M\"{o}bius symmetry
together with the one-dimensional spacetime on which it acts are encoded into
the modular inclusion of two operator algebras in a Hilbert space. A
comprehensive discussion of the relation of \textit{strongly additive} nets
via modular inclusions to M\"{o}bius covariant nets can be found in \cite{GLW}.

This abstract algebraization of the concepts behind the standard description
QFT in terms of geometrically ordered quantum matter has a generalization to
higher dimensions \cite{K-W}\cite{interface}. The number of monads is still
finite and increases with spacetime dimensions of the QFT which one wants to
construct. Since the monad has no internal structure all the physical and
mathematical richness of QFT comes from relative modular positioning of copies
of a monad in a shared Hilbert. Although in the present state of modular
operator theory QFT models (with the exception of chiral theories) cannot be
classified and constructed in this way, it illustrates an important aspect of
the holistic nature of QFT and the unexplored power of its modular
localization principle.

\section{The d=1+1 Jordan model and the isomorphism which solves the
Einstein-Jordan conundrum}

With the \textit{locally restricted vacuum} representing a highly
(non-tracial) impure state with respect to $all$ modular Hamiltonians
$H_{mod}(\mathcal{\check{O}}),$ $\mathcal{\check{O}\supseteq O~}$on local
observables $A\in\mathcal{A(O)\subset A(O}^{\prime\prime}\mathcal{)},$ a
fundamental conceptual difference of QFT and QM has been identified. In this
section it will be shown how modular localization solves the E-J conundrum in
terms of an operator-algebraic isomorphism between Jordan's model and its
Einstein statistical mechanics analog. In view of the importance of modular
localization in the present QFT research, this is much more than a historical
retrospection; it is a precursor of the Unruh-Hawking observation of an
analogy between localization-caused and heat bath thermal behavior in QFT in a
case where the thermal aspect is not only an analogy but where the two systems
(after re-parametrization) are identical. Whereas modular localization has
helped to view Unruh-Hawking situations as special cases of the defining
structural property of causal localization in QT, the full solution of the E-J
conundrum at the time of the historical dispute might have brought an aspect
of QFT into the open (just a nice dream) which may have changed the path of
history of QFT and in particular reconciled Einstein with the intrinsic
ensemble probability of QFT.

In modern terminology Jordan's model QFT of a two-dimensional
photon\footnote{This terminology was quite common in the early days of field
quantization before it was understood that that in contrast to QM, the
physical properties of QFT depend in an essential way on the spacetime
dimension. Jordan's d=1+1 "photons" and his later "neutrinos" (in his
"neutrino theory of light" \cite{Jor}) are not two-dimensional versions of
4-dimensional objects in the sense that in QM an oscillator chain always
remains the same object independent of the dimension of the embedding space.}
is really a model of a chiral current. As a two-dimensional zero mass field
which solves the wave equation it can be decomposed into its two $u,v$
lightray components (omitting the $u,v$ vector indices)%

\begin{align}
&  \partial_{\mu}\partial^{\mu}\Phi(t,x)=0,~\Phi
(t,x)=V(u)+V(v),~u=t+x,~v=t-x~\label{chiral}\\
&  j(u)=\partial_{u}V(u),~j(v)=\partial_{v}V(v),~\left\langle j(u)j(u^{\prime
})\right\rangle \sim\frac{-1}{(u-u^{\prime}-i\varepsilon)^{2}}\nonumber\\
&  ~~\ T(u)=:j^{2}(u):,~T(v)=:j^{2}(v):,~~\left[  j(u),j(v)\right]
=0~~~~~~\nonumber
\end{align}

The model belongs to the family of conformal chiral models. The scale
dimension of the chiral current is $d_{sd}(j)=1$, whereas the energy-momentum
tensor (the Wick-square of $j$) has $d_{sd}(T)=2$; the $u$ and $v$ world are
completely independent and it suffices to consider the fluctuation problem for
one chiral component. The logarithmic divergence of the zero dimensional
chiral $d_{sd}(V)=0$ current potential $V$ arises from the semiinfinite
string-localization of $V;$ a better behaved M\"{o}bius-group covariant fields
is the charge-carrying sigma model field formally written as $expi\alpha V$
\cite{BMT}. Since our argument only uses the M\"{o}bius covariance and
localization of the Weyl algebra generated by the chiral current $j,$ such
details concerning the charge superselection structure and charge-carrying
fields are irrelevant.

The E-J fluctuation problem can be formulated in terms of $j$ (charge
fluctuations) or $T$~(energy fluctuations). It is useful to recall that vacuum
expectations of chiral operators are invariant under the fractionally acting
3-parametric acting M\"{o}bius group ($x$ stands for $u,v$), which for
$j~$reads%
\begin{align}
U(a)j(x)U(a)^{\ast}  &  =j(x+a),~U(\lambda)j(x)U(\lambda)^{\ast}=\lambda
j(\lambda x)~\ ~dilation\\
U(\alpha)j(x)U(\alpha)^{\ast}  &  =\frac{1}{(-sin\alpha+cos\alpha)^{2}}%
j(\frac{cos\alpha x+sin\alpha x}{-sin\alpha x+cos\alpha x})~rotation\nonumber
\end{align}

The next step consists in identifying the KMS property of the locally
restricted vacuum with that of a global system in a thermodynamic limit state.
For obvious reasons it is referred to as the \textit{inverse Unruh effect,}
i.e. finding a heat bath thermal system which corresponds to the restriction
of the vacuum to the $j$-associated Weyl algebra $\mathcal{A}((a,b))$
localized on an interval.

\begin{theorem}
(\cite{S-W}) The global chiral operator algebra $\mathcal{A}(\mathbb{R})$
associated with the heat bath representation at temperature $\beta=2\pi$ is
isomorphic to the vacuum representation restricted to the half-line chiral
algebra such that
\begin{align}
(\mathcal{A}(\mathbb{R}),\Omega_{2\pi}) &  \cong(\mathcal{A}(\mathbb{R}%
_{+}),\Omega_{vac})\label{map}\\
(\mathcal{A}(\mathbb{R})^{\prime},\Omega_{2\pi}) &  \cong(\mathcal{A}%
(\mathbb{R}_{-}),\Omega_{vac})\nonumber
\end{align}
The isomorphism intertwines the translations of $\mathbb{R}$ with the
dilations of $\mathbb{R}_{+}$, such that the isomorphism extends to the local
algebras:
\begin{equation}
(\mathcal{A}((a,b)),\Omega_{2\pi})\cong(\mathcal{A}((e^{-2\pi a},e^{2\pi
b})),\Omega_{vac})\label{exp}%
\end{equation}

\end{theorem}

This can be shown by modular theory. The proof extends prior work by Borchers
and Yngvason \cite{Bo-Yng}. Let $\mathcal{A}$ denote the $C^{\ast}$ algebra
associated to the chiral current $j$\footnote{One can either obtain the
bounded operator algebras from the spectral decomposition of the smeared free
fields $j(f)$ or from a Weyl algebra construction.}. Consider a thermal state
$\omega$ at the (for convenience) Hawking temperature $2\pi$ associated with
the translation on the line. Let $\mathcal{M}~$be the operator algebra
obtained by the GNS representation and $\Omega_{2\pi}~$the state vector
associated to $\omega.$ We denote by $\mathcal{N}$ the halfspace algebra of
$\mathcal{M}$ and by $\mathcal{N}^{\prime}\mathcal{\cap M}~$the relative
commutant of $\mathcal{N}~in~\mathcal{M}.~$The main point is now that one can
show that the modular groups of $\mathcal{M},~\mathcal{N}$ and $\mathcal{N}%
^{\prime}\mathcal{\cap M~g~}$generate a "hidden" positive energy
representation of the M\"{o}bius group $SL(2,R)/Z_{2}$ where "hidden"
\cite{S-W} means that the actions have no geometric interpretation on the
thermal net. The positive energy representation acts on a hidden vacuum
representation for which the thermal state is now the vacuum state $\Omega
.$The relation of the previous three thermal algebras to their vacuum
counterpart is as follows:
\begin{align}
&  \mathcal{N}=\mathcal{A}(1,\infty),~\mathcal{N}^{\prime}\mathcal{\cap
M=A}(0,1),~\mathcal{M}=\mathcal{A}(0,\infty)\\
&  \mathcal{M}^{\prime}\mathcal{=A}(-\infty,0),~\mathcal{A}(-\infty
,\infty)=\mathcal{M\vee M}^{\prime}\nonumber\\
&  \mathcal{M}(a,b)=\mathcal{A}(e^{-2\pi a},e^{2\pi b})
\end{align}
Here $\mathcal{M}^{\prime}$ is the "thermal shadow world" which is hidden in
the standard Gibbs state formalism but makes its explicit appearance in the so
called \textit{thermo-field} setting i.e. the result of the GNS description in
which Gibbs states described by density matrices or the KMS stated resulting
from their thermodynamic limits are described in a vector formalism. The last
line expresses that the interval algebras are exponentially related.

In the theorem we used the more expicit notation
\[
\mathcal{M}(a,b)=(\mathcal{A}(a,b),\Omega_{th})=(\mathcal{A}(e^{-2\pi
a},e^{2\pi b}),\Omega_{vac})
\]

Moreover we see, that there is also a natural space-time structure on the
shadow world i.e. on the thermal commutant to the quasilocal algebra on which
this hidden symmetry naturally acts. Expressing this observation a more
vernacular way, one may say: \textit{the thermal shadow world has been
converted into virgin living space beyond the horizon of a localized Unruh
world } \cite{S-W}. In conclusion, we have encountered a rich hidden symmetry
lying underneath the tip of an iceberg, of which the tip was first seen by
Borchers and Yngvason \cite{Bo-Yng}.

Although we have assumed the temperature to have the Hawking value $\beta
=2\pi,$ the reader convinces himself that the derivation may easily be
generalized to arbitrary positive $\beta$ as in the Borchers-Yngvason work. A
more detailed exposition of these arguments is contained in a paper
\textit{Looking beyond the Thermal Horizon: Hidden Symmetries in Chiral Models
}\cite{S-W}.

In this way an interval of length $L$ (one-dimensional box) passes to the size
of the split distance $\varepsilon$ which plays the role of Heisenberg's
vacuum polarization cloud $\varepsilon\sim e^{-L}.$ Equating the thermodynamic
$L\rightarrow\infty$ with the the limit of a fuzzy localization converging
against a sharp localization on the vacuum side in $(e^{-2\pi L},~e^{2\pi L})$
for $L\rightarrow\infty$ with the fuzzynes $e^{-2\pi L}\equiv\varepsilon
\rightarrow0,~$the thermodynamic limit of the thermal entropy passes to that
of the localization entropy in the limit of vanishing $\varepsilon$%
\begin{equation}
LkT\mid_{kT=2\pi}\simeq-\text{ln}\varepsilon
\end{equation}
where the left hand side is proportional to the (dimensionless) heat bath
entropy and the right hand side is proportional to the localization entropy.

Although it is unlikely that a localization-caused thermal system is generally
isomorphic to a heat bath thermal situation in higher dimensions (the strong
inverse Unruh effect), there may exist a "weak" inverse Unruh situation in
which the volume factor corresponds to a logarithmically modified
dimensionless area law (previous section) i.e. $(\frac{R}{\Delta R})^{n-2}%
$ln$(\frac{R}{\Delta R})~$instead of $(\frac{R}{\Delta R})^{n-2}$ where $R$ is
the radius of a double cone with a fuzzy surface and $\frac{\Delta R}{R}$ the
dimensionless measure of the fuzzy surface. The box on the localization side
(\ref{st}) has two transverse- and one lightlike- extension and is the
counterpart of the spatial box in a weak inverse Unruh picture. This would be
different by a logarithmic factor from the area law which is suggested by the
analogy to the behavior of vacuum polarization of a partial charge in the
sharp localization limit (Section 2) and which also appears in 't Hooft's
"brickwall" proposal \cite{brickwall} to make the derivation of the Hawking
radiation of quantum matter in CST consistent with Bekenstein's classical area
law. The present state of computational control of the split property is not
able to decide between these two possibilities.

The above isomorphism shows that Jordan's imagined situation of quantum
fluctuations interpreted as fluctuations in a small subinterval of a chiral
QFT restricted to an interval which is M\"{o}bius equivalent to a halfline and
therefore isomorphic to Einstein's thermodynamic limit system on the full
line. Although the thermal aspect of a restricted vacuum in QFT is a
structural consequence of causal localization, the general identification of
the dimensionless modular temperature with a temperature of a heat bath system
(the inverse Unruh effect), or, which is equivalent, the modular "time" with
the physical time, is an unsolved conceptual-mathematical problem \cite{Con}.

\section{The role of modular localization in the ongoing research}

The use of the Einstein-Jordan conundrum as an indicator of the presence of
radically different aspects of QFT would remain in the philosophic-historical
realm if the LQP algebraic viewpoint would not also lead to new results beyond
Lagrangian quantization and perturbation theory. In fact these new local
algebraic methods led, for the first time in the history of QFT, to an
existence proof in the presence of interactions in a family of d=1+1 models
with realistic (non-canonical) short distance behavior\footnote{The previous
existence proofs were limited to models with a canonical (superrenormalizable)
short distance behavior \cite{Gl-Ja}.}. This is the content of pathbreaking
work by Gandalf Lechner \cite{Lech}\cite{Le}. In this way the the existence of
certain integrable models about which there were already exact computational
results for certain formfactors \cite{Kar} was finally secured and a new
direction for future more general attempts was pointed out. What is most
surprising is the radically new method of construction. Whereas the Lagrangian
(or functional integration) quantization starts from a classical action and
the result of the perturbative calculation reveals its quantum interpretation
("bottom-to-top"), in the LQP approach to QFT one starts from foundational
principles and tries to find the appropriate mathematical concepts to
implement them (top-to-bottom).

Bottom-to-top methods aim typically at perturbative series for correlation
functions of fields ("off-shell"); particle aspects ("on-shell") as scattering
amplitudes and formfactors usually appear at a later stage. Despite all its
merits concerning observational agreements, the well-known divergence of
perturbative series limits its use; mathematically-controlled approximations
are not known, let alone proofs of existence of interacting models. Since
these problems are endemic, they could have their explanation in the direct
use of rather singular objects (operator-valued distributions). Top-to-bottom
approaches typically start from on-shell quantities as the $S_{scat}$-matrix
or formfactors (interacting operators between outgoing bra and incoming ket
states) whereas off-shell correlation functions only appear in later stages.
The important observation is that in QFTs with a complete particle
interpretation the $S_{scat}$-matrix is a relative modular invariant of the
wedge-localized algebra (\ref{scat}). This suggests that the position of a
wedge algebra in an interating theory is determined in terms of the $S_{scat}%
$-matrix and the corresponding free field wedge algebra.

This idea can be explicitly tested for a particular class of d=1+1 theories
whose $S_{scat}$ is determined in terms of multi-component two-particle
scattering functions fulfilling the Yang-Baxter relation. It has been known
for a long time that elastic scattering solutions of the so-called S-matrix
bootstrap setting (Poincar\'{e} invariance, unitarity and crossing) cannot
exist in higher dimensions; the necessary presence of inelastic processes
prevents the constructive use of such general properties. Elastic $S_{scat}%
$-matrices in d=1+1 are however susceptible to a classification within a
bootstrap setting. Such a classification in terms of symmetries leads to
families of models which only in a few cases make perturbative contact with
Lagrangians. The second step, namely the construction of formfactors of a
would-be QFT associated with these scattering functions known as the
"bootstrap-formfactor project", has led to a wealth of results \cite{Z}.

QFT is the only area of theoretical physics which, apart from certain d=1+1
models with canonical (superrenormalizable) short distance behavior
\cite{Gl-Ja}, remained in its almost 90 years history without mathematical
support concerning the existence of interesting interacting models. The
"factorizing models", with their nontrivial renormalizable short distance
behavior and their integrable formfactors associated with elastic S-matrices,
present a fascinating "theoretical laboratory" for the study of this ultimate
conceptual challenge, namely to secure the mathematical existence and (in case
of non-integrable models) provide mathematically controlled approximations.

This last step in this construction is the verification that the computed
formfactors really fulfill all the properties which entered in form of an
Ansatz into their construction. This is the most interesting and important
part of the construction since it includes the new LQP idea of proving
existence of a family of models with non-canonical short distance behavior
\cite{Lech}. Unlike the Lagrangian quantization setting, ultraviolet and
renormalization aspects play no role in this construction. In a few cases a
perturbative comparison of formfactors with those obtained from Lagrangian
quantization identified them with Lagrangian models whose integrability was
based on quasiclassical arguments (Sine-Gordon, Sinh-Gordon,..); but for most
of the integrable model it was necessary to baptize them with names referring
to symmetries or analogies with lattice models.

The first observations about a relation between these "bootstrap-formfactor
methods" with localized operator algebras and their generators go back the 90s
\cite{S2} and consisted in attributing a spacetime interpretation to the
Zamolodchikov-Faddeev algebra generators which the Zamolodchikov brothers
introduced as a simplifying algebraic device for the classification of
factorizing $S_{scat}$-matrices. It turned out that the Fourier transforms of
these generalized creation/annihilation operators are generators of
wedge-localized algebras. The existence proof in d=1+1 requires an even more
subtle step: to show that a nontrivial double cone localized algebra can be
obtained from intersecting a wedge-localized algebra $\mathcal{A}(W)$ with a
translate of its opposite$~\mathcal{A}(W^{\prime}).~$This was achieved by the
use of modular nuclearity in a pathbreaking work by Lechner \cite{Lech}. These
ideas have been extended by deformation theory (deformation of free fields for
models without bound states \cite{Le}), and meanwhile integrable models which
even by experts are considered to be difficult (as the $O(N)$-model
\cite{O(N)}) are in the range of the modular nuclearity arguments \cite{L-S}
which already secured the existence of simpler models.

In contrast to Wightman's setting of QFT in terms of correlation functions of
fields, Haag's formulation of a QFT in terms of local nets of operator
algebras fortunately (since this seems to be prohibitively difficult) does not
require an explicit construction of singular fields (operator-valued
distributions) since all observationally important quantities (S-matrix,
formfactors) can be computed in the LQP setting. Factorizing models and chiral
models \cite{Ka-Lo} are presently the only QFTs for which the operator-algebra
based methods led to existence proofs. These are examples par excelence for
what is meant by "top-to-bottom calculations".

The algebraic method reveals much more than a new construction scheme of
certain integrable models. It also leads to a foundational understanding of
the crossing property and an intrinsic distinction between integrable and
non-integrable (the typical case) based on properties of generators of
wedge-localized operator algebras. The important observation is that
one-particle vacuum-polarization-free-generators (PFG) of wedge-localized
algebras and their generalized multi-particle "emulats" come with two
distinctively different operator properties \cite{BBS}: either their domains
are translational invariant and they permit a Fourier-transformation or their
domains are only invariant under those transformations which leave the wedge
invariant. In the first case one can show that either $S_{scat}=1$ or (and
this requires d=1+1) $S_{scat}$ is elastic\footnote{In fact it can be shown
that there are no higher connected elastic S-matrices than $S_{scat}^{(2)}$ so
that the n-particle amplitude is a combinatorial product of two-particle
amplitudes.}, whereas in the second case one has to cope with the full
$S_{scat}$ which couples the two-particle state to all higher particle states
within the same superselection sector.

This situation permits a very elegant and useful definition of integrability
in QFT which bypasses the (in QFT unhandy) definition in terms of infinitely
many conserved charges in involution.

\begin{definition}
(\cite{integrable}) The dichotomy integrable/nonintegrable in QFT is defined
in terms of temperate/nontemperate PFG generators of wedge-localized algebras.
Temperate generators with $S_{scat}\neq1$ only exist in d=1+1
\end{definition}

It is very informative to present the definition of these two types of
operators. In both cases they are bijectively related to wedge-localized
incoming fields which share with the interacting algebras the same
representation of the Poincar\'{e} group. Since therefore the modular
unitaries $\Delta^{it}$ (the W-preserving Lorentz boost) for all wedge
algebras with the same $\mathcal{P}$-representration coalesce, the domains
$domS$ of the different Tomita $S$-operators agree and hence all states
$\left\vert \eta\right\rangle $ in $domS~$permit a representation $A\left\vert
0\right\rangle $ in terms of different operators, each uniquely affiliated
with one of the different algebras. Our interest is to relate a specific
interacting wedge-localized algebra $\mathcal{A}(W)$ with the corresponding
interaction-free incoming algebra $\mathcal{A}_{in}(W).$

Using the notation $A(f)$ for a free field smeared with a test function $f$
with $suppf\in W$ and denoting the bijective related operator affiliated to
$\mathcal{A}(W)$ by $\left(  A(f)\right)  _{\mathcal{A}(W)}$ we have%

\begin{align}
&  A(f)\left\vert 0\right\rangle =A(f)_{\mathcal{A}(W)}\left\vert
0\right\rangle ,~A(f)=A(\check{f})\label{bij}\\
&  \mathcal{A}_{in}(W)\ni A\longleftrightarrow A_{\mathcal{A}(W)}%
\in\mathcal{A}(W)\nonumber\\
&  (A_{\mathcal{A}(W)})^{\ast}\left\vert 0\right\rangle =SA_{\mathcal{A}%
(W)}\left\vert 0\right\rangle =S_{scat}A^{\ast}\left\vert 0\right\rangle
\nonumber
\end{align}
where the $\in$ for unbounded operators means "affiliated with". $W$-smeared
free fields $A(f)$ may also be written in terms of their momentum space
creation/annihilation operators integrated with on-shell wave functions
$\check{f}(p),$ which are the mass-shell projection of the Fourier-transformed
test functions $f.$ The $suppf\subset W~$property implies that $\check{f}$ is
a boundary value of a function which is analytic in the $(0,i\pi)$ strip of
the $W$-associated rapidity variable $\theta,~$where the lower boundary
corresponds to the physical one-particle wave function $\check{f}$ and the
upper boundary value is its complex conjugate (the c. c. of the anti-particle
in case of complex fields). The third line in (\ref{bij}) states that the
bijection does not preserve the passing to the adjoint and at the same time
introduces the $S_{scat}$-matrix of the specific interacting model.

Only in the temperate (integrable) case the Fourier transformed $\left(
\tilde{A}(p)\right)  _{\mathcal{A}(W)}$ exist \cite{BBS}. In this case the PFG
behave very much like Wightman fields i.e. the $\left(  A(\check{f})\right)
_{\mathcal{A}(W)}$ exist for all wave functions (equivalently for all Schwartz
smearing function). The only difference is its localization; it is in a
certain restricted sense a nonlocal object: for $suppf\supset W$ it is fully
nonlocal and for $suppf\subset W$ it remains wedge-localized independent of
whether one sharpens this localization to $suppf\subset\mathcal{O}\subset W.$

It turns out that d=1+1 temperate one-particle PFG fulfill the Z-F algebra
commutation relations which for the simplest case of a meromorphic scattering
function $S(\theta)$ (no Yang-Baxter structure) reads:%
\begin{align}
&  \left(  \tilde{A}(\theta)\right)  _{\mathcal{A}(W)}\equiv Z^{\ast}%
(\theta),~Z^{\ast}(\theta-i\pi):=Z(\theta)\\
Z(\theta)Z(\theta^{\prime}) &  =\delta(\theta-\theta^{\prime}+i\pi
)+S(\theta-\theta^{\prime})Z(\theta^{\prime})Z(\theta),\ S(-\theta
)=\overline{S(\theta)}=S(\theta+i\pi)\nonumber
\end{align}
where the definition in the first line is just a notation which permits to
write the various commutation relations between creation/annihilation
components in terms of one formula (second line) so that the $\delta$-function
only contributes to the mixed $Z$-$Z^{\ast}~$commutation relations. The
physical scattering range in $S(\theta)~$is $\theta>0$ and the relations for
the meromorphic function $S(\theta)$ in the second line represent unitarity
and crossing.

For later purposes it is necessary to generalize the one-particle PFGs in
terms of multiparticle "emulats". With $A(f_{1},..f_{n})\equiv:A(f_{1}%
)..A(f_{n}):,~suppf_{i}\subset W~~$the same modular arguments based on the
domain properties of the Tomita $S$ \cite{BBS} which led to the wedge
localized PFGs also apply to the emulats
\[
A(f_{1},..f_{n})_{\mathcal{A}(W)}\left\vert 0\right\rangle =A(f_{1}%
,..f_{n})\left\vert 0\right\rangle =\left\vert \check{f}_{1},..\check{f}%
_{n}\right\rangle _{in}%
\]
As in the single particle case the unique affiliation of a multiparticle
emulate with the interacting algebra $\mathcal{A}(W)$ is secured by its
appropriately defined commutation with the commutant $\mathcal{A}(W)^{\prime
}.\,$The Wick-ordering simplifies the connection with n-particle states. Even
for temperate emulats the bijection does not respect the algebraic
multiplication structure i.e. $(A(f)A(g))_{\mathcal{A}(W)}\neq
A(f)_{\mathcal{A}(W)}A(g)_{\mathcal{A}(W)}.$

For the derivation of the crossing identity for formfactors one needs the idea
of \textit{analytic ordering changes} of $W$-associated rapidities. Consider
the formfactor which describes the vacuum-polarization components of a local
excitation$~$in d=1+1$.$ The n-particle components of a bra state $B^{\ast
}\left\vert 0\right\rangle $ is a symmetric function in the rapidities. The
degeneracy of statistics may be encoded into an ordering prescription;
assuming bosonic statistics we write
\begin{equation}
\left\langle 0\left\vert B\right\vert \theta_{1}....\theta_{n}\right\rangle
,~B\in\mathcal{A}(W),~~if~\theta_{1}>\theta_{2}>..>\theta_{n}%
\end{equation}
The ordering referes to the numerical values of the $\theta~$and not to the
ordering of there indices. Other orderings are interpreted as the result of an
analytic change of the $\theta s.$ For theories with meromorphic scattering
functions $S(\theta)$ the rapidity is a uniformization variable for the
formfactors, i.e. the property of being meromorphic is passed to the
formfactors. In this case an analytic transposition of two adjacent $\theta$
is given by the multiplication of the ordered formfactor by the scattering
functions $S(\theta_{i}-\theta_{i+1})$ and repeated transpositions generate a
non-degenerate "analytic" representation of the permutation group \cite{BKFZ}
and the Z-F algebra restricted to creation operators is an algebraic encoding
of these analytic transpositions.

The crossing identity will be shown to result from the cyclic KMS identity
which is a consequence of the modular localization property of $\mathcal{A}%
(W).$ We need it in the form%

\begin{align}
&  \left\langle 0|B\left(  A^{(1)}\right)  _{\mathcal{A}(W)}\left(
A^{(2)}\right)  _{\mathcal{A}(W)}|0\right\rangle =\left\langle 0|\left(
A^{(2)}\right)  _{\mathcal{A}(W)}\Delta B\left(  A^{(1)}\right)
_{\mathcal{A}(W)}|0\right\rangle \label{K}\\
&  A^{(1)}\equiv:A(f_{1},..f_{k}):,~A^{(2)}\equiv:A(f_{k+1},..f_{n}%
):,~suppf_{i}\in W,~B\in\mathcal{A}(W)\nonumber
\end{align}
where two of the $\mathcal{A}(W)$ operators are emulates. Whenever an emulat
acts on the vacuum it creates a W-localized multi-particle state ($\left\vert
\hat{f}^{(a)}\right\rangle \equiv A(f)^{\ast}\left\vert 0\right\rangle $):
\begin{align}
&  \left\langle 0|B\left(  A(\check{f}_{1},..\check{f}_{k})\right)
_{\mathcal{A}(W)}|\hat{f}_{k+1},..\hat{f}_{n}\right\rangle _{in}%
=~_{out}\left\langle \hat{f}_{k+1}^{(a)},..\hat{f}_{n}^{(a)}|\Delta B|\hat
{f}_{1},..\hat{f}_{k}\right\rangle _{in}\equiv~~\label{1}\\
&  \equiv\int d\theta_{1}..\int d\theta_{n}~\hat{f}_{1}(\theta),..\hat{f}%
_{n}(\theta)_{out}\left\langle \bar{\theta}_{k+1},.\bar{\theta}_{n}%
|\Delta^{\frac{1}{2}}B|\theta_{1},.\theta_{k}\right\rangle _{in}\nonumber
\end{align}
The bra states on the right side refer to antiparticles and the second line
results from analytical continuation by $-i\pi$ of the complex conjugate
antiparticle wave functions which are equal to the original wave functions, so
that on both sides of the identity in the first line (\ref{1}) the dense set
of wave functions agrees. In order to write the left hand side in terms of an
n-particle-vacuum formfactor, we need to know how the emulat act on a
k-particle state. If it would be possible to extend this relation from the
dense set in the space of $L^{2}(\theta)$-integrable wave function to wave
functions with compact support in $\theta~$our analytic ordering assumption
suggests the identification%
\begin{align}
&  \left\langle 0|B\left(  A(\check{f}_{1},..\check{f}_{k})\right)
_{\mathcal{A}(W)}|\check{f}_{k+1},..\check{f}_{n}\right\rangle _{in}%
=\left\langle 0|B|\check{f}_{1},....\check{f}_{n}\right\rangle _{in}\\
&  for~\ \ \ supp\check{f}_{1}>...>supp\check{f}_{n}\nonumber
\end{align}
whatever is the result of analytic reorderings may be. In fact since the
statistics of the states and the symmetry inside a Wickproduct always permit
to order inside the state and inside the emulat, it is only the relative
ordering of the emulat cluster with respect to the particle cluster in the
state which matters. This results in the particle crossing relation
\begin{align}
&  \left\langle 0|B|\theta_{1},..\theta_{n}\right\rangle _{in}=~_{out}%
\left\langle \theta_{k+1},..\theta_{n}|\Delta^{\frac{1}{2}}B|\theta
_{1},..\theta_{k}\right\rangle _{in}\equiv\\
&  \equiv~_{out}\left\langle \theta_{k+1}+i\pi,..\theta_{n}+i\pi|B|\theta
_{1},..\theta_{k}\right\rangle _{in},~~(\theta_{1},..\theta_{k})>(\theta
_{k+1},..\theta_{n})\nonumber
\end{align}
In other words the crossing identity is an extended form of the KMS identity;
whereas the particle wave functions in the KMS relation are analytic (coming
from W-localzed test functions), the crossing identity is a formfactor
identity which only requires a relative ordering between bra and ket clusters
but no integration with analytic wave functions. Rewritten interms of
$p$-variables one recovers the standard form%
\begin{equation}
\left\langle 0|B|p_{1},..p_{n}\right\rangle _{in}=~_{out}\left\langle
-p_{k+1},..-p_{n}|B|p_{1},..p_{k}\right\rangle _{in} \label{cro}%
\end{equation}
Since temperate PFG are synonymous with integrability, the assumptions made
about analytic ordering changes and their relation to the action of emulats on
particle states are established ex post facto using the exactly computed formfactors.

The conceptual and the calculational situation gets much more complicated in
case of non-integrable QFT. Assuming again that the action of an emulat on a
mutiparticle state can be related to an analytic change from an ordered
situation, it is clear that this cannot be reduced to subsequent analytic
transpositions. In other words the reordering in the second line%
\begin{equation}
\left\langle 0\right\vert B(\tilde{A}(\theta))_{\mathcal{A}(W)}\left\vert
\theta_{1},..\theta_{n}\right\rangle =%
\genfrac{\{}{.}{0pt}{}{\left\langle 0\right\vert B\left\vert \theta,\theta
_{1},..\theta_{n}\right\rangle ,~\theta>\theta_{1}>..>\theta_{n}%
}{\sum_{\left\vert \ddddot{\vartheta}\right\vert }\int d\ddddot{\vartheta
}F(\ddddot{\vartheta};\theta_{1},.\theta_{k};\theta)\left\langle 0\right\vert
B\left\vert \ddddot{\vartheta},\theta,\theta_{k+1},..\theta_{n}\right\rangle }%
\end{equation}
which is necessary if $...>\theta_{k}>\theta>$ $\theta_{k+1}>..$has to be done
in one sweep. A single term in the resulting expression is an integral of a
multiparticle state of total particle number $\left\vert \ddddot{\vartheta
}\right\vert +1+n-k$ integrated with a function $F$ over the multivariable
configuration $\ddddot{\vartheta}\equiv(\vartheta_{1},..\vartheta_{m})$ of
m=$\left\vert \ddddot{\vartheta}\right\vert $ particles. For the functions $F$
there exists an Ansatz in terms of a "grazing shot amplitude" which can be
written in terms of the full scattering amplitudes \cite{integrable}. It
expresses the fact that there is no direct interaction between the $\theta
_{i}$ i.e. the $\theta~$must pass through the ($\theta_{1},..\theta_{k}$)
cluster for activating an interaction. Hence the result of re-establishing the
total ordering including $\theta$ leaves the configuration with smaller
$\theta_{i}$-values than $\theta$ unchanged, but cause a particle-number
non-preserving change of the swept cluster.

From the exact results in \cite{BBS} one knows that $(\tilde{A}(\theta
))_{\mathcal{A}(W)}$ cannot make sense as operators. The tacit assumption
underlying the above Ansatz is that such objects exist as bilinear forms
between multiparticle states (formfactors of $(\tilde{A}(\theta))_{\mathcal{A}%
(W)}$) and that the action of operators $A(\check{f})$ on localized
multiparticle domains can subsequently be constructed from the knowledge of
the bilinear forms. The generalization from one-particle PFGs to multiparticle
emulats appears to require a clever notation more than new concepts.

The best way to place this attempt to get a hold on interacting QFT with a
complete particle interpretation into a historical context is to view it as an
extension of Wigner's representation theoretical particle setting (which by
the use of modular wave function localization passes functorially to
interaction free local operator algebras (section 3)) to the realm of
interactions when the second quantization functor has to be replaced by
emulation \cite{integrable}.

Explicit formulas for bilinear forms $(\tilde{A}(\theta))_{\mathcal{A}(W)}$
can be found in \cite{integrable}. Although they pass the consistency check of
reducing to the corresponding much simpler action of temperate PFGs, what
remains to be done is to show that one can obtain operators $\left(
A_{in}(f)\right)  _{\mathcal{A}(W)}$ with the claimed domain properties and,
last not least, that these operators are wedge dual in the sense%

\begin{equation}
\left\langle \psi\left\vert \left[  JA(\hat{f})_{\mathcal{A}(W)}J,A(\hat
{g})_{\mathcal{A}(W)}\right]  \right\vert \varphi\right\rangle =0,~J=S_{scat}%
J_{in} \label{weloc}%
\end{equation}
which is the wedge duality expressed in terms of the emulats. These problems
are very difficult for non-integrable models, so that one does not expect an
(positive or negative) answer in the near future.

The derivation of the crossing identity in the non-integrable case follows the
same line of reasoning as for integrable models. Although the vacuum
formfactors for non-integrable models are not meromorphic functions, it is
plausible that their singularities are accounted for by the multiparticle
scattering threshold (cuts from multiple roots) which the $\theta
$-uniformization cannot remove. In that case the denseness of the analytic
wave function can be used as before (to separate the $\theta$-ordered from the
rest). This leads again to the "kinematical" crossing formula (\ref{cro}%
)\ which fortunately does not require any knowledge about how emulats depend
on the interaction. From the crossing relation of formfactors one can derive
the pair crossing of scattering amplitudes using LSZ reduction formulas.

The crossing of the elastic scattering amplitude was derived in a tour de
force based on the use of the theory of multivariable analytic functions in
\cite{BEG}. The ordering limitation of the crossing identity which seems to
have no counterpart in the formal LSZ derivation is in reality a property
which exactly matches the non-overlapping limitations caused by hitting
threshold singularities in the derivation of the LSZ reduction formalism from
the rigorous Haag-Ruelle scattering theory \cite{Bu-Su}.

Besides the nonperturbative new insights modular localization also led to
ideas to use the mild short distance behavior of string-localized fields which
allows couplings within the power-counting limit (the prerequisite for
renormalizability) for \textit{all spins.} In particular for s=1
string-localized potentials resolve the clash between localization and the
Hilbert space structure of (m=0,s=1) representations by sacrificing the
point-localization which is not only conceptually more reasonable than the
BRST approach (which abandons the Hilbert space in favor of Krein spaces) but
also permits to investigate problems which are not accessible to the BRST
setting \cite{integrable}. These problems are presently under intense
investigations \cite{MSY2}.

The bridge between the Einstein-Jordan conundrum as the oldest
Gedankenexperiment pointing towards consequences of modular localization and
issues on the frontier of particle theory has been a source of great
fascination to the author.

\textit{Note added}: Thanks to John Stachel I became aware of an interesting
and most comprehensive article on the Unruh effect by John Earman \cite{Ear}.
For a philosopher of science the interesting question is the concrete
realization in terms of existing (or Gedanken-) observable hardware of Haag's
quantum adaptation of the classical causal localization principle which is
much more subtle than its intuitive support (finite spatial extension, finite
duration of counter activation) which led him to present QFT in the LQP
setting of local algebras. The present article on the other hand is concerned
with the complete solution of the E-J conundrum in the concrete context of
Jordan's chiral model (d=1+1 "photon") as an isomorphism between a restricted
vacuum state (Jordan) with a global heat-bath system (Einstein). According to
my best knowledge this model is the only known QFT which realises what has
been called the "inverse Unruh effect" \cite{S-W}. Concerning Hawking
radiation, it is my firm conviction that the understanding of formation of
black holes (which involves ideas which are not covered by modular
localization \cite{F-H}) is more important than finding a global state on the
Kruskal extension which plays the same role for the localized region outside
the Schwarzschild horizon as a restricted Minkowski vacuum. The modular
thermalization is a property of the ensemble of all observables which are
modular localized within the same region in the sense of Haag's LQP; but as in
global statistical mechanics, the KMS condition is also satisfied by each
individual operator of the localized ensemble. It is certainly not suitable as
an egg-boiling device \cite{Ear} since everything in such a causally closed
world, including the "modular cook" will be boiled. Since a modular
localization in $\mathcal{O}~$is also localized in every $\widehat
{\mathcal{O}}$ $\supset\mathcal{O},$ the $\mathcal{O}$-localized operators
fulfill a continuous set of KMS relations with different modular Hamiltonians.
This modular "tightness", which is totally absent in QM, shows that QFT is
rightfully considered as a foundational QT; the prize to be paid for this
enormous conceptual distance to QM is that all mathematically controlled
approximation methods of the latter (notably single operator methods based on
spectral resolution of selfadjoint operators) are powerless in QFT, and
functional integral representations, which is shared between both, do not
permit a mathematical control since the perturbative series diverges in QFT.

Of special interest to the author have been those manifestations of KMS
properties of modular localization which do not require to think about a
relation of the modular temperature to the one measured in terms of a
thermometer as e.g. the (possibly logarithmically modified)
\textit{proportionality of the localization entropy to the dimensionless area
}$A/\varepsilon^{2}.$ It is rather improbable that Bekenstein's black hole
entropy formula can be related to the vacuum polarization contribution near
event horizons which are connected with the Hawking radiation (quantum matter
in CST); more plausible is that it refers to the contribution of gravity
degrees of freedom in a future QGR. Another problem is the concept of
information loss. Information theory has its conceptual home in QM where
entanglement related to the decomposition of pure states with respect to a
factorization into two subsystems leads to impure states in terms of
averaging. The restriction of the QFT vacuum (more generally finite energy
states) to a local subsystem obtained from modular localization however does
\textit{not need any averaging} over the causal complement. It is questionable
whether information theory can be generalized to such situations.

The most rewarding results of wedge-localization is however the understanding
of the formfactor \textit{particle crossing} as a relic of the KMS cyclicity.
The $\iota\pi$-strip analyticity needed for the return to physical formfactors
together with the change from incoming ket particles to outgoing bra states
corresponds to the KMS cyclicity together with the analyticity encoded in the
"modular" temperature $T_{mod}=1/2\pi.$ This includes also the modular
construction recipe for integrable models and the future promise to get a
nonperturbative constructive hold on realistic models of QFT in a new
top-to-bottom approach. The Unruh effect is certainly part of a bigger story
about a structural property of QFT which is independent on such fleeting
observer-dependent effects and doubts about relations of the modular
temperature. In \cite{Solveen} it is shown that, although Unruh's correlations
are thermal, the \textit{local} temperature at a point fluctuates wildly
around $T_{average}=0$\footnote{In view of the results in \cite{Solveen} where
it is shown that the KMS $\beta$ has in general (in particular in non-inertial
systems and in the presence of curvature) no relation to the temperature in
the sense of the zeroth thermodynamic law (local equilibrium), }$.$

\textbf{Acknowledgement}: I am indebted to Jens Mund for discussions at the
start of this project. I also thank Detlev Buchholz for reading the manuscript
and suggesting some changes in the last section.


\begin{thebibliography}{99}                                                                                               %


\bibitem {E1}A. Einstein, Physikalische Zeitschrift \textbf{10}, (1909) 185

\bibitem {E2}A. Einstein, Physikalische Zeitschrift \textbf{18}, (1917), 121

\bibitem {Thesis}P. Jordan, Zeitschrift f\"{u}r Physik \textbf{30} (1924) 297

\bibitem {E3}A. Einstein, \textit{Bemerkungen zu P. Jordans: Zur Theorie der
Quantenstrahlung}, Zeitschrift f\"{u}r Physik \textbf{30}, (1925) 784

\bibitem {Du-Ja}A. Duncan and M. Janssen, \textit{Pascual Jordan's resolution
of the conundrum of the wave-particle duality of light}, arXiv:0709.3812

\bibitem {BHJ}M. Born, W. Heisenberg and P. Jordan, \textit{Zur
Quantenmechanik II}, Zeitschr. f\"{u}r Physik \textbf{35}, (1926) 557

\bibitem {E-G}H. Epstein and V. Glaser, The role of locality in perturbation
theory. Ann. Inst. H. Poincar\'{e} A\textbf{19} (1973), 211

\bibitem {Haag}R. Haag, \textit{Local Quantum Physics}, Springer 1996

\bibitem {S1}B. Schroer, B. Schroer, Modular localization and the
bootstrap-formfactor program, Nucl. Phys. \textbf{B499}, 1997, 547, hep-th/9702145

\bibitem {S2}B. Schroer, \textit{Modular localization and the d=1+1 formfactor
program}, Annals of Physics \textbf{295}, (1999) 190-223

\bibitem {Born}M. Born, \textit{Quantum mechanics in impact processes},
Zeitschrift f\"{u}r Physik \textbf{38}, (1926) 803-840

\bibitem {Coe}F. Coester and W. N. Polyzou, Phys. Rev. D \textbf{26}, (1982)
1348-1379 and references therein

\bibitem {interface}B. Schroer, \textit{Studies in History and Philosophy of
Modern Physics }\textbf{41} (2010) 104--127, arXiv:0912.2874

\bibitem {Lille}R. Haag, \textit{Discussion of the `axioms' and the asymptotic
properties of a local field theory with composite particles }(historical
document), Eur. Phys. J. H 35, 243--253 (2010)

\bibitem {1934}W. Heisenberg, Verhandlungen der S\"{a}chsischen Akademie der
Wissenschaften zu Leipzig, \textbf{86}, (1934) 317-322

\bibitem {Wight}R. F. Streater and A. S. Wightman, PCT, \textit{Spin and
Statistics and all that}, New York, Benjamin 1964

\bibitem {Co}A. Connes, Ann. Inst. Fourier \textbf{24}, (1974) 121

\bibitem {Con}A. Connes and C. Rovelli, \textit{Von Neumann Algebra
Automorphisms and Time-Thermodynamics Relation in General Covariant Quantum
Theories}, Quant. Grav.\textbf{11},(1994) 2899

\bibitem {Do-Lo}S. Doplicher and R. Longo, Inv. math. \textbf{75}, (1984) 493

\bibitem {Jakob}J. Yngvason, \textit{The role of type III factors in quantum
field theory}, Reports in Mathematical Physics \textbf{55}, (2005) 135-147, arXiv:math-ph/0411058

\bibitem {Bi-Wi}J.\ J. Bisognano and E. H. Wichmann, \textit{On the duality
condition for quantum fields}, Journal of Mathematical Physics \textbf{17},
(1976) 303

\bibitem {Mu}J. Mund, \textit{An Algebraic Jost-Schroer Theorem for Massive
Theories}, arXiv:1012.1454

\bibitem {BGL}R. Brunetti, D. Guido and R. Longo, \textit{Modular localization
and Wigner particles}, Rev. Math. Phys. \textbf{14}, (2002) 759

\bibitem {S-W}B. Schroer and H-J Wiesbrock, Rev. Math. Phys. 12 (2000) 461-473

\bibitem {Unruh}W. G. Unruh, \textit{Notes on black hole evaporation}, Phys.
Rev. \textbf{D14}, (1976) 870-892

\bibitem {Sew}G. L. Sewell, Annals of Physics \textbf{141}, (1982) 201

\bibitem {hol}B. Schroer, \textit{The holistic structure of causal quantum
theory, its implementation in the Einstein-Jordan conundrum and its violation
in more recent particle theories}, arXiv:1107.1374

\bibitem {K-W}R. Kaehler and H.-P. Wiesbrock, \textit{Modular theory and the
reconstruction of four-dimensional quantum field theories}, Journal of
Mathematical Physics \textbf{42}, \ (2001) 74-86

\bibitem {Do-Ro}S. Doplicher and J. E. Roberts, \textit{Why there is a field
algebra with a compact gauge group describing the superselection structure in
particle physics}, Commun. Math. Phys. \textbf{131}, (1990) 51-107

\bibitem {Ho-Wa}S. Hollands and R. M. Wald, General Relativity and Gravitation
\textbf{36}, (2004) 2595-2603

\bibitem {Jost}R. Jost: \textit{TCP-Invarianz der Streumatrix und
interpolierende Felder}, Helvetica Phys. Acta \textbf{36}, (1963) 77

\bibitem {Schweber}S. S. Schweber, \textit{QED and the men who made it; Dyson,
Feynman, Schwinger and Tomonaga}, Princeton University Press 1994

\bibitem {Dar}O. Darrigol, \textit{The origin of quantized matter fields},
Historical Studies in the Physical and Biological Sciences \textbf{16} (1986) 198-253

\bibitem {brickwall}G. 't Hooft, Int. J. Mod. Phys. A\textbf{11,} (1996) 4623

\bibitem {Robinson}O. Bratteli and D. W. Robinson, \textit{Operator Algebras
and Quantum Statistical Mechanics I and II}, Springer-Verlag Berlin Heidelberg
New York 1987

\bibitem {Sw}B. Schroer, \textit{particle physics in the 60s and 70s and the
legacy of contributions \ by J. A. Swieca}, Eur. Phys. J.H. \textbf{35},
(2010) 53, arXiv:0712.0371

\bibitem {Requ}M. Requardt, Commun. Math. Phys. \textbf{50}, (1976) 256

\bibitem {integrable}B. Schroer, \textit{The foundational origin of
integrability in quantum field theory}, arXiv:1109.1212

\bibitem {nonloc}B. Schroer, \textit{An alternative to the gauge theoretic
setting}, Found. Phys. \textbf{41}, 1543-1568,2011, arXiv:1107.1374

\bibitem {BMS}B. Schroer, \textit{Bondi-Metzner-Sachs symmetry, holography on
null-surfaces and area proportionality of \textquotedblleft light-slice"
entropy}, Foundations of Physics \textbf{41}, (2011) 204, \ arXiv:0905.4435

\bibitem {Summers}S. J. Summers, \textit{Yet More Ado About Nothing: The
Remarkable Relativistic Vacuum State}, arXiv:0802.1854

\bibitem {Kharkov}P. Jordan, \textit{The Present State of Quantum
Electrodynamics}, in \textit{Talks and Discussions of the Theoretical-Physical
Conference in Kharkov} (May 19.-25., 1929) Physik.Zeitschr.XXX, (1929) 700

\bibitem {MSY}J. Mund, B. Schroer and J. Yngvason, Commun. Math. Phys.
\textbf{268}, (2006) 621

\bibitem {Yng}M. Plaschke and J. Yngvason, \textit{Massless, String Localized
Quantum Fields for Any Helicity}, arXiv:1111.5164

\bibitem {H-S}R. Haag and B. Schroer, J. Math. Phys. \textbf{3}, (1962) 248

\bibitem {Hal}H. Halvorson, \textit{Reeh-Schlieder defeats Newton-Wigner: on
alternative localization schemes in quantum field theory}, Philosophy of
Science \ \textbf{68}, (2001) 111-133

\bibitem {HHW}R. Haag, N. M. Hugenholtz and M. Winnink, Commun. Math. Phys.
\textbf{5}, (1967) 215

\bibitem {Bo}H-J. Borchers, \textit{On revolutionizing quantum field theory
with Tomita's modular theory}, J. Math. Phys. \textbf{41}, (2000) 8604

\bibitem {Fa-Sc}L. Fassarella and B. Schroer, \textit{Wigner particle theory
and local quantum physics}, J. Phys. A \textbf{35}, (2002) 9123-9164

\bibitem {quarter}B. Schroer, Int. J. of Mod. Phys. B6, 2041 (1992)

\bibitem {Wies}H.-W. Wiesbrock, Commun. Math. Phys., \textbf{158}, (1993) 537

\bibitem {GLW}D. Guido, R. Longo and H.-W. Wiesbrock,
Commun.Math.Phys.\textbf{192}, (1998) 217

\bibitem {BMT}D. Buchholz, G. Mack and I. Todorov, Localized automorphisms for
the U(1) current, in \textit{Algebraic Theory of Superselection Sectors},
edited by D. Kastler, (World SciemtificSingapor) 1990.

\bibitem {char}B. Schroer, \textit{An alternative to the gauge theory
setting}, Foun. of Phys. \textbf{41}, (2011) 1543, arXiv:1012.0013 arXiv:1006.3543

\bibitem {Jor}B. Schroer, \textit{Pascual Jordan's legacy and the ongoing
research in quantum field theory}, arXiv:1010.4431

\bibitem {Bo-Yng}H-J. Borchers and J. Yngvason, J. Math. Phys. \textbf{40}
(1999) 601

\bibitem {foun}B, Schroer, Found. Phys. \textbf{40}, 2010, 1800, arXiv:0905.4006

\bibitem {Lech}G. Lechner, \textit{On the Construction of Quantum Field
Theories with Factorizing S-Matrices}, PhD thesis, arXiv:math-ph/0611050

\bibitem {Kar}H. Babujian and M. Karowski, Int. J. Mod. Phys. \textbf{A1952},
(2004) 34, \ and references therein to the beginnings of the
bootstrap-formfactor program

\bibitem {Z}A. B. Zamolodchikov and A. Zamolodchikov, AOP \textbf{120}, (1979) 253

\bibitem {Gl-Ja}J. Glimm and A. Jaffe, \textit{Boson quantum field theory
models, in mathematics of contemporary physics}, edited by R. F. Streater
(Academic press, London) 1972

\bibitem {Ka-Lo}Y. Kawahigashi and R. Longo, Ann. of Math.\textbf{160} (2004), math-ph/0201015

\bibitem {O(N)}H. Babujian, A. Foerster and M. Karowski, \textit{The Nested
Off-shell Bethe ansatz and O(N) Matrix Difference Equations,} arXiv:1204.3479

\bibitem {L-S}G. Lechner and Ch. Sch\"{u}tzenhofer, \textit{Towards an
operator-algebraic construction of integrable global gauge theories}, arXiv:1208.2366

\bibitem {BBS}H. J. Borchers, D. Buchholz and B. Schroer,
\textit{Polarization-free generators and the S-matrix}, Communun. in Math.
Phys. \textbf{219} (2001) 125-140

\bibitem {BKFZ}H. Babujian, A. Fring, M. Karowski and A. Zapletal, Nucl. Phys.
\textbf{B 583} (FS) 535

\bibitem {BEG}J. Bros, H. Epstein and V. Glaser, Com. Math. Phys. \textbf{1},
(1965) 240

\bibitem {Bu-Su}D. Buchholz and S. J. Summers, \textit{Scattering in
Relativistic Quantum Field Theory: Fundamental Concepts and Tools}, arXiv:math-ph/0509047

\bibitem {MSY2}joint project of J. Mund, B. Schroer and J. Yngvason, work in progress

\bibitem {F-H}K. Fredenhagen and R. Haag, Commun. Math. Phys. \textbf{127},
(1990) 273

\bibitem {Ear}J. Earman, \textit{The Unruh effect for philosophers}, SHPMP
\textbf{42}, (2011) 81

\bibitem {Le}G. Lechner, \textit{Deformations of quantum field theories and
integrable models}, arXiv:1104.1948

\bibitem {Solveen}Ch. Solveen, \textit{Local Thermal Equilibrium and KMS
states in Curved Spacetime}, arXiv:1211.0431
\end{thebibliography}
\end{document}